\documentclass[12pt]{article}
\pdfoutput=1

\usepackage{draft} 
\usepackage{hyperref}
\usepackage{graphicx,color,subfig}
\usepackage{multirow}
\usepackage{cite}
\usepackage{mciteplus}
\usepackage{skak}
\usepackage{empheq}
\DeclareFontFamily{OT1}{pzc}{}
\DeclareFontShape{OT1}{pzc}{m}{it}{<-> s * [1.10] pzcmi7t}{}
\DeclareMathAlphabet{\mathpzc}{OT1}{pzc}{m}{it}

\def\be#1\ee{\begin{align}#1\end{align}}

\begin{document}

\unitlength = .8mm

\begin{titlepage}

\begin{center}

\hfill \\
\hfill \\
\vskip 1cm

\title{Hamiltonian Truncation Study of Supersymmetric Quantum Mechanics: S-Matrix and Metastable States}

\author{Bruno Balthazar, Victor A. Rodriguez,  Xi Yin}

\address{
Jefferson Physical Laboratory, Harvard University, \\
Cambridge, MA 02138 USA
}

\email{bbalthazar@g.harvard.edu, victorrodriguez@g.harvard.edu, xiyin@fas.harvard.edu}


\end{center}

\abstract{We implement the Rayleigh-Ritz method in supersymmetric quantum mechanics with flat directions, and extract the S-matrix and metastable resonances. The effectiveness of the method is demonstrated in two strongly coupled systems: an ${\cal N}=1$ toy supermembrane model, and an ${\cal N}=4$ model with a $U(1)$ gauge multiplet and a charged chiral multiplet.}

\vfill

\end{titlepage}

\eject

\tableofcontents

\section{Introduction} 

Supersymmetric quantum mechanics (SQM) \cite{Witten:1981nf, Cooper:1994eh} have played important roles in elucidating the structure of vacua in quantum field theories \cite{Witten:1982df, Witten:1982im}, the spectrum of solitons \cite{Sen:1994yi, Sethi:1995zm, Denef:2002ru, Alim:2011ae}, and in holography \cite{Callan:1996dv, Banks:1996vh, Maldacena:1997re, Banks:1997hz, Balasubramanian:1997kd, Susskind:1998vk, Polchinski:1999ry, BrittoPacumio:1999ax}. While much is known about the structure of ground states in SQM \cite{Witten:1982df, Yi:1997eg, Sethi:1997pa, Moore:1998et, Frohlich:1999zf, Lin:2014wka, Cordova:2014oxa, Hori:2014tda, Cordova:2015qka, Cordova:2015zra}, the study of dynamics of excited states has been largely limited to perturbation theory \cite{Becker:1997cp, Becker:1997xw, Plefka:1997hm, Taylor:2001vb} and thermodynamics based on preliminary field theoretic methods \cite{Kabat:1999hp, Iizuka:2001cw, Smilga:2008bt, Lin:2013jra} and Monte Carlo simulation \cite{Anagnostopoulos:2007fw, Hanada:2009ne, Hanada:2016zxj}. The much less understood real time dynamics of strongly coupled SQM, on the other hand, are of utmost interest: in principle, they capture unitary quantum evolution of black hole microstates through holographic dualities.

The Rayleigh-Ritz method, or the Hamiltonian truncation method, has been widely applied to the spectral problem of bounded quantum mechanical systems \cite{reed1978iv}, and to the study of renormalization group flows in strongly coupled quantum field theories \cite{Yurov:1989yu, Hogervorst:2014rta, Rychkov:2014eea, Rychkov:2015vap}. In this paper, we develop this method to analyze the S-matrix and metastable states in SQM. While the method itself does not rely on supersymmetry, the latter provides a natural class of models with flat directions and rich dynamics. We will demonstrate the effectiveness of the method in two nontrivial strongly coupled models: (1) the ${\cal N}=1$ toy supermembrane model \cite{deWit:1988xki, Frohlich:1999zf}, and (2) ${\cal N}=4$ SQM with a $U(1)$ gauge multiplet coupled to a charged chiral multiplet \cite{Denef:2002ru, Anous:2015xah}. The method is implemented through the following steps.

\noindent 1. We separate the Hamiltonian into a free part $H_0$ and an interaction potential $V$, and introduce an IR cutoff at a sufficiently large distance $L$.

\noindent 2. We work with a basis of wave functions that diagonalize the free Hamiltonian $H_0$ in a ``box" of size $L$, and truncate the basis to a finite set by keeping eigenfunctions up to truncation energy $\Lambda$. We then evaluate the matrix elements of the full Hamiltonian $H_0+V$ on this truncated basis, and diagonalize it numerically to find the energy levels. The separation of free and interaction Hamiltonian is such that the interaction potential $V$ is sufficiently smooth, which leads to small mixing of low frequency modes with high frequency modes, and fast convergence of the energy levels with increasing $\Lambda$.

\noindent 3. The energy levels $E_n$ of the truncated Hamiltonian for a scattering spectrum are dense in the limit of large $L$. We subtract from the number of states $n$ up to energy $E_n$ a universal IR contribution governed by the asymptotic scattering wave function, and obtain a ``renormalized number of states" $\overline n(E_n)$. We refer the collection of points $(E_n, \overline{n}(E_n))$ as the {\it spectral set}.

\noindent 4. We collect the spectral sets for different and sufficiently large values of IR cutoff $L$. Provided that the truncation energy $\Lambda$ is sufficiently high, over a finite range of energy $E$ of interest, we will find that the spectral set lies on the union of $k$ smooth curves, $k$ being the number of asymptotic regions, or the effective dimension of the S-matrix at a given energy. In the $k=1$ case, where the S-matrix is a single scattering phase, the curve traced out by the spectral set determines the scattering phase $\phi(E)$ as a function of energy. In the $k>1$ case, by varying the details of the IR cutoff on different asymptotic regions, one can determine the full $U(k)$ S-matrix, up to a relative $U(1)^{k-1}$ phase ambiguity of the asymptotic scattering wave functions.

\noindent 5. A metastable state corresponds to a jump of the scattering phase $\phi(E)$ by $2\pi$. The spectrum of metastable states and their decay width are determined from the peaks of $d\phi(E)/dE$. Our procedure also allows for the explicit determination of the wave function of metastable resonances.

Our method is explained in more detail in the next section. The convergence with truncation energy is discussed through one dimensional examples in section \ref{warmup}. The application to the toy supermembrane model and the ${\cal N}=4$ SQM will be presented in section \ref{toy} and \ref{sqm} respectively. We conclude with some prospectives on the Rayleigh-Ritz approach to holographic models in section \ref{discussion}.

\section{S-matrix and the density of states}
\label{general}


Let us consider a quantum mechanical system with $k$ one-dimensional asymptotic regions where the dispersion relation takes the form $E=p^2$, $p$ being the asymptotic momentum. The in and out states will be denoted $|E,i\rangle^{in}$ and $|E,i\rangle^{out}$ respectively, $i=1,\ldots, k$. They are related by the S-matrix
\ie
|E,i\rangle^{in} = \sum_j S_{ij}(E) |E,j\rangle^{out},
\fe
where $S_{ij}(E)$ is an $k\times k$ unitary matrix. An asymptotic wave function takes the form
\ie
\sum_i \left( a_i e^{-ip x} + e^{ipx} \sum_j a_jS_{ji}\right) |i\rangle.
\fe
We now introduce an IR cutoff by placing hard walls at distance $x=L_i$ in the $i$-th asymptotic region, so that the spectrum is discretized. The quantization condition
\ie
a_i e^{-ip L_i} + e^{ipL_i} \sum_j a_j S_{ji}(E)=0
\fe
amounts to demanding that the $k\times k$ matrix
\ie
B_{ij}(E) = e^{-ipL_i} \delta_{ij} + e^{ipL_i} S_{ji}(E)
\fe
admits a zero eigenvalue, or equivalently, $\det B(E)=0$.

In the simplest $k=1$ case, the S-matrix is a single scattering phase $e^{i\phi(E)}$. The asymptotic quantization condition is
\ie\label{phie}
\phi(E_n) + 2L\sqrt{E_n} = 2\pi (n+{1\over 2}),~~~n=0,1,2,\ldots.
\fe
Given the spectrum $\{E_n\}$, we can extract the scattering phase from 
\ie
\phi(E) = 2\pi \overline n(E) ,
\fe
where the renormalized number of states is given by
\ie
\overline n(E) = \lim_{L\to \infty} \left[n(E)+{1\over 2} - \frac{L\sqrt{E}}{\pi}\right].
\fe
In practice, we only need to take $L$ to be greater than the effective range of interaction, and collect the spectral set
\ie\label{elset}
\left\{ \Big( E_n, 2\pi (n+{1\over 2}) - 2L \sqrt{E_n} \Big), ~~~n=0,1,2,\ldots\right\}
\fe
for a sequence of values of $L$, all of which lies on the graph of $\phi(E)$.

If there is a metastable state of energy $E_*$ and decay width $\epsilon$, we expect the scattering phase to behave as
\ie
e^{i\phi(E)} \sim e^{i\phi_0} {E-E_*- i\epsilon\over E-E_* + i\epsilon}
\fe
for $E$ close to $E_*$. This leads to a peak in the derivative of the scattering phase,
\ie
{d\phi(E)\over dE} \sim {2\epsilon\over (E-E_*)^2 + \epsilon^2}.
\label{eq:BW}
\fe

Now let us generalize this prescription to the case of several asymptotic regions, namely $k>1$. To begin with, take $L_i = L$, and denote by $e^{i\phi_j(E)}$ ($j=1,\ldots,k$) the eigenvalues of $S_{ij}(E)$. The asymptotic quantization condition can be written as
\ie
\phi_{i_n}(E_n) + 2L\sqrt{E_n} = 2\pi (n+{1\over 2}),~~~n=0,1,2,\ldots,~~~i_n \in \{1,\ldots,k\}.
\label{eq:aqc2}
\fe
In other words, now the spectral set (\ref{elset}) lies on the {\it union of $k$ curves} that are the graphs of the functions $\phi_i(E)$, $i=1,\ldots,k$. Numerically, it is again useful to combine the sets (\ref{elset}) for different values of $L$ (provided that they are larger than the effective interaction range).

A slight modification of this prescription allows for extracting the full $U(k)$ S-matrix, up to conjugation by a diagonal unitary matrix (since the choice of phase for each asymptotic wave function is a priori ambiguous). Let us take $L_i = L+y_i$, with $y_i$ finite while taking the large $L$ limit. Now the set of points (\ref{elset}) lie on the $k$ curves defined by the $k$ eigenvalues of the matrix
\ie{}
\widehat S_{ij}(E;y) = e^{i y_i \sqrt{E}} S_{ij}(E) e^{i y_j \sqrt{E}}
\fe
as functions of $E$.

As a nontrivial example, consider the $k=2$ case. Up to conjugation by a diagonal $U(2)$ matrix, we can write the S-matrix as
\ie
S = e^{i {\phi_1+\phi_2\over 2}} e^{{i\over 2} \sigma_3 \A}e^{i\sigma_2 \theta} e^{{i\over 2} \sigma_3 \A},
\fe 
where $\phi_1+\phi_2\over 2$ is the overall phase, and $\A(E)$ and $\theta(E)$ are to be determined from the spectral set (\ref{elset}). Now take $y_1=y/2$, $y_2=-y/2$, so that 
\ie
\widehat S = e^{{i\over 2}\sigma_3 y \sqrt{E}} S e^{{i\over 2}\sigma_3 y \sqrt{E}}.
\fe
The eigenvalues of $\widehat S$ are $\exp\left[ i {\phi_1(E)+\phi_2(E)\over 2} \pm i \gamma(E;y) \right]$, with
\ie
\cos\gamma(E;y) = \cos\theta(E) \cos(\A(E) + y\sqrt{E}\,).
\fe
$\gamma(E;y)$ is determined from the difference between the two curves traced out by the spectral set.
By maximizing $|\cos \gamma(E;y)|$ with respect to $y$, we then determine both $\theta(E)$ and $\A(E)$.

\section{Quantum scattering in one dimension from Rayleigh-Ritz method: dependence on truncation energy}
\label{warmup}

To gain some intuition, let us consider the scattering problem in one dimension, with the Hamiltonian $H=-\partial_x^2+V(x)$ for $x>0$, where the potential $V(x)$ vanishes sufficiently fast in the $x\to \infty$ limit. We also impose the boundary condition that the wave function vanishes at $x=0$. In the Hamiltonian truncation approach, we place a hard wall at $x=L$, and work with the truncated basis of standing waves in the empty box of length $L$, 
\ie\label{odb}
\psi_n(x) = \sqrt{2\over L}\sin{n\pi x\over L},~~~~n=1,2,\ldots,N.
\fe
We shall choose $N$ such that the energy at the truncation level 
\ie
\Lambda = \left(N\pi\over L \right)^2
\fe
is much bigger than the scattering energy $E$. The matrix elements of the Hamiltonian on this basis are given by
\ie\label{hnm}
H_{nm}\equiv \langle\psi_n |H|\psi_m\rangle = \left( {n\pi\over L} \right)^2 \delta_{nm} + {2\over L} \int_0^L dx V(x) \sin {n \pi x\over L} \sin {m \pi x\over L} .
\fe
One then proceeds to diagonalize the $N\times N$ truncated Hamiltonian matrix $H_{(N)} \equiv ( H_{nm})_{1\leq n,m\leq N}$, and read off the scattering $\phi(E)$ from the renormalized number of states up to energy $E$ as in (\ref{phie}).

The efficiency of the Hamiltonian truncation method relies on the convergence rate of the eigenvalues of the truncated Hamiltonian $H_{(N)}$ with increasing $N$ or $\Lambda$. This may be estimated from (\ref{hnm}) in the limit of large $n$ and fixed $m$. The off-diagonal entries of $H_{nm}$ are given by the Fourier transform of the function $V(x)$, extended to an odd function over the range $x\in [-L,L]$. Assuming the latter has bounded $k$-th order derivative in $x$, the off-diagonal $H_{nm}$ decays with $n$ at least as fast as $(n/L)^{-k-1}$, and its contribution to the eigenvalues of the Hamiltonian is bounded by $(n/L)^{-2k-4}$. The error due to truncation at $N$ may be estimated by summing over $n\geq N$, giving a result that scales like $N^{-2k-3}\sim \Lambda^{-k-{3\over 2}}$.
If the potential $V(x)$ is smooth, we expect the eigenvalues of the truncated Hamiltonian converge with increasing truncation energy $\Lambda$ faster than any inverse power of $\Lambda$.

\begin{figure}
\centering
\subfloat[]{\includegraphics[width=5.5cm]{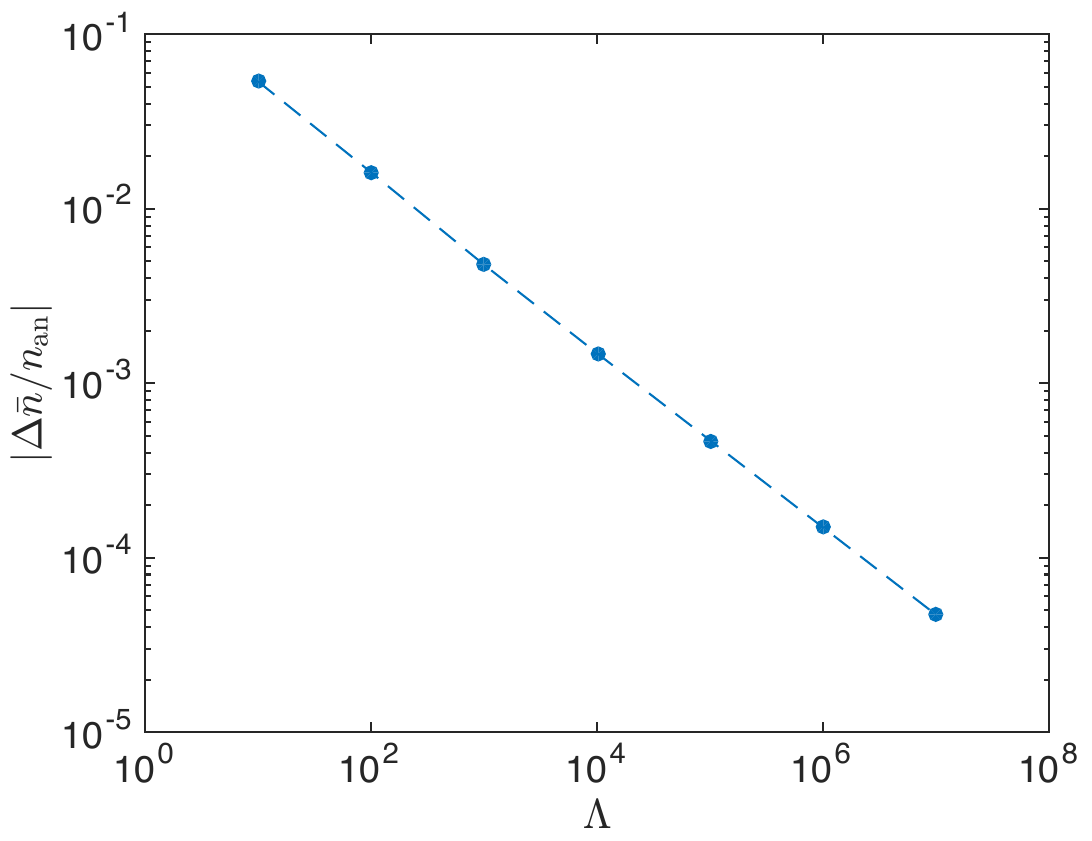}}
\subfloat[]{\includegraphics[width=5.5cm]{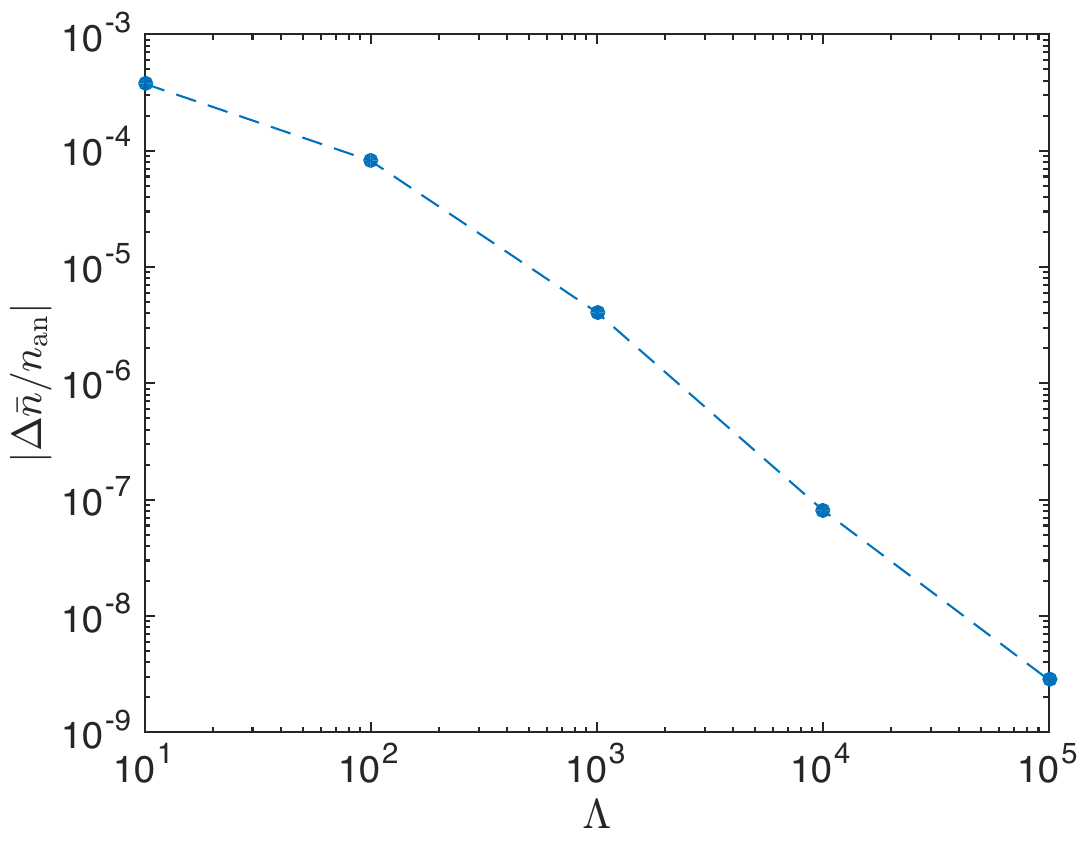}}
\subfloat[]{\includegraphics[width=5.5cm]{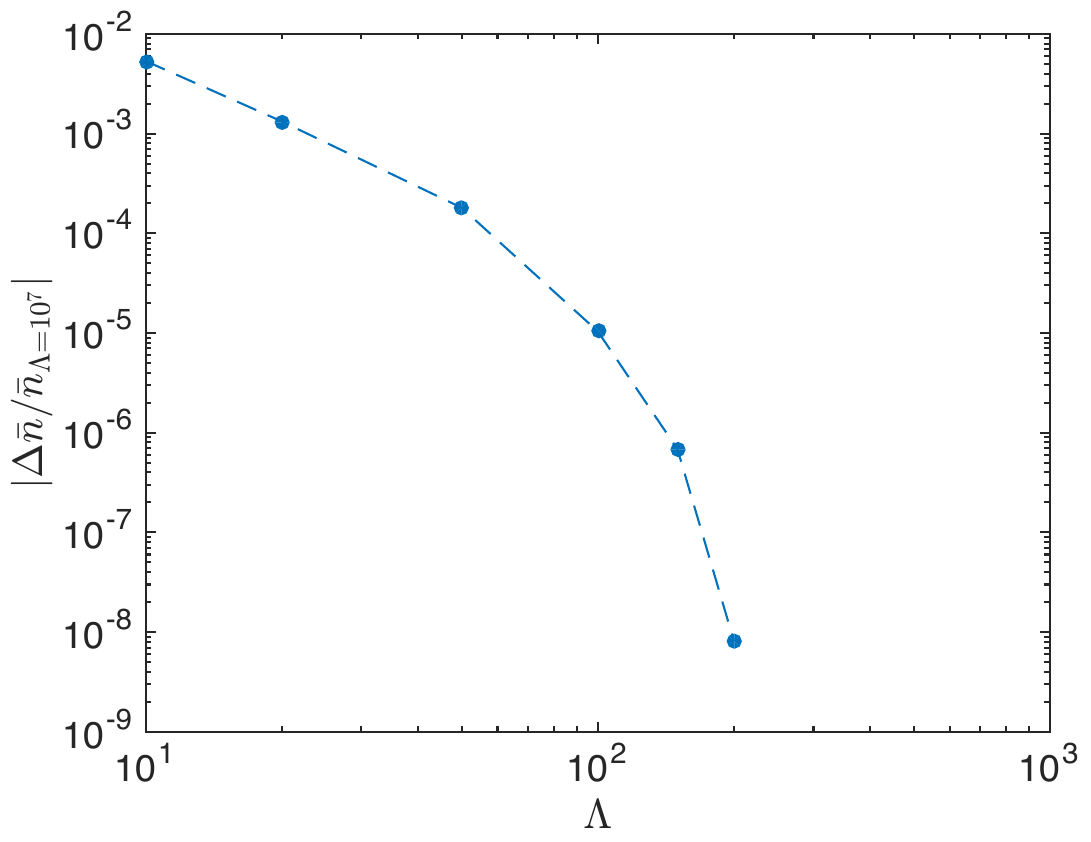}}
\caption{Log-log plot of error in the density of states $\overline{n}(E)$ due to a finite truncation energy $\Lambda$, with fixed IR cutoff $L=10$, in three examples: (a) scattering in delta function potential $V(x)=\delta(x-1)$ at energy $E\simeq2.6$; (b) scattering in a rectangular potential barrier of height 1 in the range $x\in [0.95,1.05]$, at energy $E\simeq1.6$; (c) scattering in the presence of smooth potential barrier $V=x^2 e^{-10 (x-1)^2}$, at energy $E\simeq2.5$. In (a), (b), the error is computed by comparison with the analytic result, whereas in (c) the error is computed by comparison with the result at very high truncation energy $\Lambda = 10^7$. }
\label{fig:1derr}
\end{figure}

As an example, consider the delta function potential,
\ie
& V(x) = c \delta(x-a).
\label{eq:deltapot}
\fe
The matrix elements of the Hamiltonian on the basis (\ref{odb}) are given by
\ie
H_{nm} = \left( {n\pi\over L} \right)^2 \delta_{nm} + {2c\over L} \sin {n \pi a\over L} \sin {m \pi a\over L} .
\fe
The eigenvalues of the truncated Hamiltonian $H_{(N)}$ are given by solutions of
\ie
f_N(E)\equiv 1 + {2c\over L} \sum_{n=1}^N {\sin^2({\pi a n \over L})\over ({\pi n\over L})^2-E} = 0.
\fe
In this case, the error due to the level truncation can be estimated from
\ie
|f_N(E) - f_\infty(E)| < {2c\over \pi\sqrt{E}} {\rm arctanh} \sqrt{E\over\Lambda} \approx {2c\over \pi \sqrt{\Lambda}} ~~(\Lambda\gg E).
\fe
The error in the energy levels is suppressed by $\sqrt{E\over \Lambda}$, agreeing with the expectation of $\Lambda^{-k-{3\over 2}}$ for $k=-1$. This is also confirmed by numerics in Figure \ref{fig:1derr}(a).

If we consider a rectangular potential $V(x)$, which is bounded but its first order derivative in $x$ unbounded, the energy levels are expected to converge with truncation energy as $\Lambda^{-{3\over 2}}$. The numerical results are shown in Figure \ref{fig:1derr}(b). On the other hand, if the potential $V(x)$ is smooth, we expect faster-than-inverse-power convergence with the truncation energy. An example of the form $V(x) = c x^2 \exp[-b(x-a)^2]$ is considered in Figure \ref{fig:1derr}(c), and indeed the convergence with truncation energy $\Lambda$ appears to be exponential.


%

The lesson here is that, in order to apply the Hamiltonian truncation method efficiently, we should split the Hamiltonian into a ``free" part and an interaction ``potential", $H=H_0 + V$, such that while the basis functions diagonalize $H_0$, the interaction potential $V$ is sufficiently smooth so that its matrix elements $V_{nm}$ decays sufficiently fast in the $n\to \infty$ limit for fixed $m$.

\section{The toy supermembrane model}
\label{toy}

The toy supermembrane model, introduced in \cite{deWit:1988xki} (see also \cite{Frohlich:1999zf}), is one of the simplest nontrivial SQM that admits flat directions and a gapless spectrum of scattering states. It is based on a Hilbert space of two-component wave functions in two variables $x, y$, with a single Hermitian supercharge $Q$, given by
\ie
Q = i \partial_x \sigma_3 + i\partial_y \sigma_1 - xy \sigma_2 
\fe
The Hamiltonian is 
\ie
H = Q^2 = -\partial_x^2 - \partial_y^2 + x^2y^2 + x\sigma_3 - y \sigma_1.
\fe
The model admits a discrete symmetry of the dihedral group, which is represented projectively by the generators
\ie
& U = {\bf P}_x \sigma_1, ~~~~ V = {\bf P}_y \sigma_3, ~~~~ W = {\bf R} e^{\pi i \sigma_2/4},
\fe
where ${\bf P}_x$ and ${\bf P}_y$ are parity transform in $x$ and $y$ respectively, and ${\bf R}$ is the rotation by $\pi\over 2$ on the plane: $x\mapsto y$, $y\mapsto -x$. They obey $U^2=V^2=-W^4=1$, $UWU=W^{-1}$. $U$ and $V$ commute with the supercharge $Q$, whereas $W^{-1}QW=-Q$.

\subsection{The scattering problem}

For a given energy $E>0$, there are four scattering in-states $|E, i\rangle^{in}$, and four out-states $|E,i\rangle^{out}$, related by
\ie
|E, i\rangle^{in} = S_{ij}(E) |E, j\rangle^{out}
\fe
The index $i=1,\ldots,4$ label the four asymptotic regions $x\to \mp\infty$ ($y\to 0$), and $y\to \mp\infty$ ($x\to0$). The asymptotic states form a representation $R$ of the dihedral symmetry group. Explicitly, the asymptotic scattering wave function of $|E,1\rangle^{in}$ takes the form
\ie
\langle x,y|E,1\rangle^{in} \sim {1\choose 0} \left( e^{- i\sqrt{E}|x|} + S_{11} e^{ i \sqrt{E} |x|} \right) e^{-|x| y^2/2} |x|^{1\over 4} ,~~~x\to -\infty,
\fe
and has only outgoing waves in the other three asymptotic regions, namely $x\to +\infty$ and $y\to \pm \infty$. Likewise, the asymptotic wave functions of $|E,i\rangle^{in}$, $i=2,3,4$, obey
\ie
& \langle x,y|E,2\rangle^{in} \sim {0\choose 1} \left( e^{- i\sqrt{E}|x|} + S_{22} e^{ i \sqrt{E} |x|} \right) e^{-|x| y^2/2} |x|^{1\over 4} ,~~~x\to \infty,
\\
& \langle x,y|E,3\rangle^{in} \sim {{1\over \sqrt{2}}\choose -{1\over \sqrt{2}}} \left( e^{- i\sqrt{E}|y|} + S_{33} e^{ i \sqrt{E} |y|} \right)   e^{-|y| x^2/2} |y|^{1\over 4} ,~~~y\to -\infty,
\\
& \langle x,y|E,4\rangle^{in} \sim {{1\over \sqrt{2}}\choose {1\over \sqrt{2}}} \left( e^{- i\sqrt{E}|y|} + S_{44} e^{ i \sqrt{E} |y|} \right) e^{-|y| x^2/2} |y|^{1\over 4} ,~~~y\to \infty,
\fe
and each of them has only outgoing waves in the other three asymptotic regions. The asymptotic wave functions for the out states can be constructed similarly.
The dihedral symmetry generators are represented on these asymptotic states by the $4\times 4$ matrices
\ie
\rho(U) = \begin{pmatrix} 0 & 1 & 0 & 0 \\ 1 & 0 & 0 & 0 \\ 0 & 0 & -1 & 0 \\ 0 & 0 & 0 & 1 \end{pmatrix},~~~ \rho(V) = \begin{pmatrix} 1 & 0 & 0 & 0 \\ 0 & -1 & 0 & 0 \\ 0 & 0 & 0 & 1 \\ 0 & 0 & 1 & 0 \end{pmatrix},~~~
\rho(W) = \begin{pmatrix} 0 & 0 & 0 & 1 \\ 0 & 0 & -1 & 0 \\ 1 & 0 & 0 & 0 \\ 0 & 1 & 0 & 0 \end{pmatrix}.
\fe
The supercharge $Q$ is represented by a matrix $\rho^{in}(Q)$ on the basis of in-states and by $\rho^{out}(Q)$ on the basis of out-states, where
\ie
\rho^{in}(Q) = - \rho^{out}(Q) = \sqrt{E} \begin{pmatrix} -1 & 0 & 0 & 0 \\ 0 & -1 & 0 & 0 \\ 0 & 0 & 1 & 0 \\ 0 & 0 & 0 & 1 \end{pmatrix}.
\fe
The S-matrix lies in a singlet representation contained in $R^*\otimes R$. Further demanding time reversal symmetry and supersymmetry fixes $S(E)$ up to a phase, namely
\ie
S(E) = e^{i\phi(E)}  \begin{pmatrix} 0 & 0 & {1\over \sqrt{2}} & {1\over \sqrt{2}} \\ 0 & 0 & -{1\over \sqrt{2}} & {1\over \sqrt{2}} \\ {1\over \sqrt{2}} & -{1\over \sqrt{2}} & 0 & 0 \\ {1\over \sqrt{2}} & {1\over \sqrt{2}} & 0 & 0 \end{pmatrix}.
\fe

\subsection{Hamiltonian truncation}

To proceed, we place hard walls at $x=\pm L$ and at $y=\pm L$, and work with the basis of wave functions
\ie
f_{n,m}(x,y) = {1\over L} \sin{n\pi (x-L)\over 2L} \sin{m\pi (y-L)\over 2L},~~~n,m\geq 1
\fe
for each component of $\psi(x,y)$. We can simplify our task slightly by restricting to the $U=+1$ sector, which amounts to working with the basis
\ie
{} & \psi_{n,m}(x,y) = {1\over L} \sin{n\pi (x-L)\over 2L} \sin{m\pi (y-L)\over 2L} {{1\over \sqrt{2}}\choose {1\over \sqrt{2}} },~~~n~{\rm even},
\\
{} & \psi_{n,m}(x,y) = {1\over L} \sin{n\pi (x-L)\over 2L} \sin{m\pi (y-L)\over 2L} {{1\over \sqrt{2}}\choose -{1\over \sqrt{2}} },~~~n~{\rm odd}.
\fe
We then numerically diagonalize the matrix
\ie
\langle n,m|H|k,\ell\rangle &= \int_{-L}^L dx \int_{-L}^L dy \, \psi_{n,m}^\dagger(x,y) H \psi_{k,\ell}(x,y)
\\
&= \delta_{nk}\delta_{m\ell} {\pi^2(n^2+m^2)\over 4L^2} + \int_{-L}^L dx \int_{-L}^L dy \, \psi_{n,m}^\dagger(x,y) \left( x^2y^2+x\sigma_3-y \sigma_1 \right) \psi_{k,\ell}(x,y).
\fe

\begin{figure}
\centering
\subfloat{\includegraphics[width=8cm]{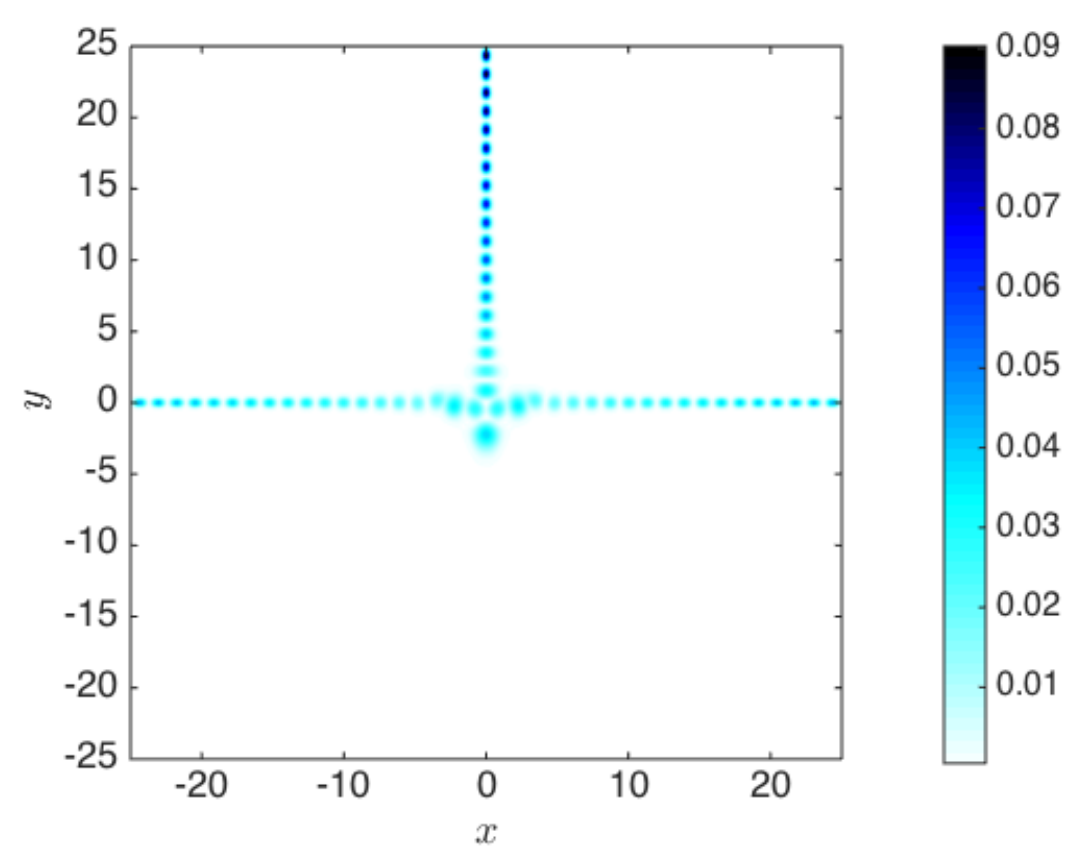}}
\subfloat{\includegraphics[width=8cm]{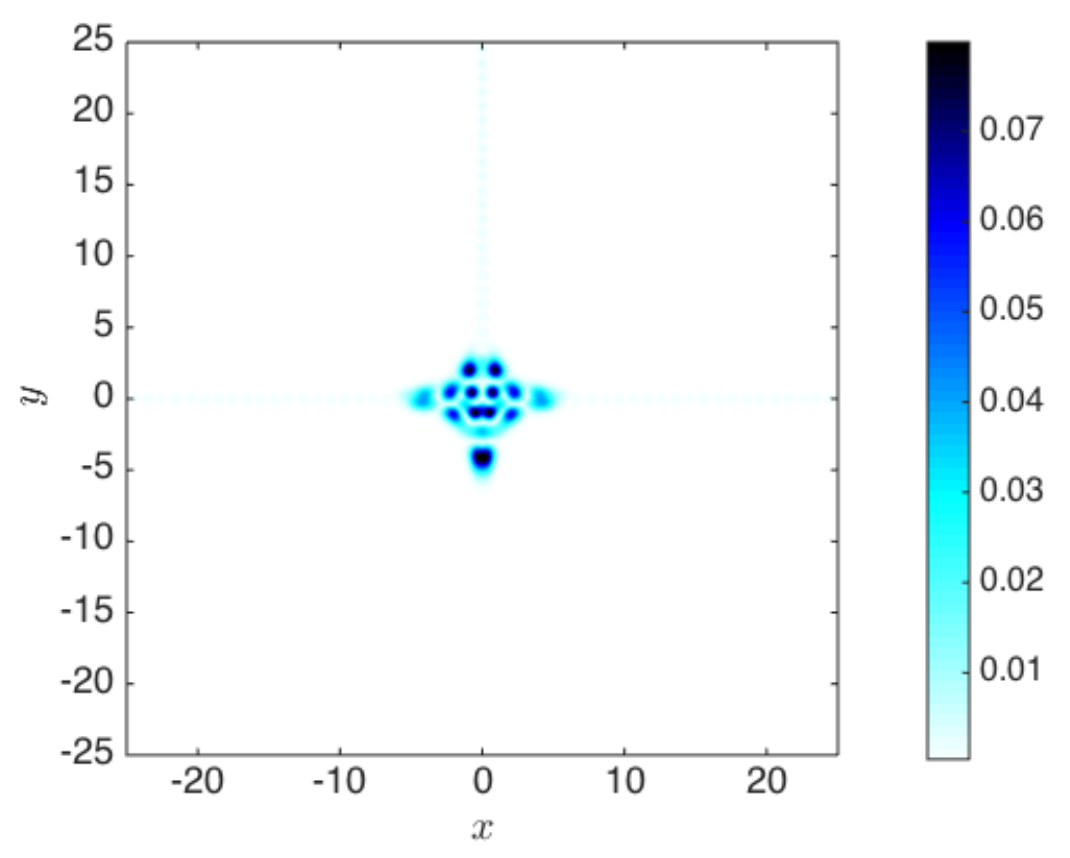}}
\caption{Probability density of a generic asymptotic wave function (left) and of a metastable wave function (right) with energies $E=5.82$ and $E=10.05$ respectively, where we have taken $L=25$, $\Lambda=380$.}
\label{fig:toywavefns}
\bigskip
\bigskip
\subfloat{\includegraphics[width=7cm]{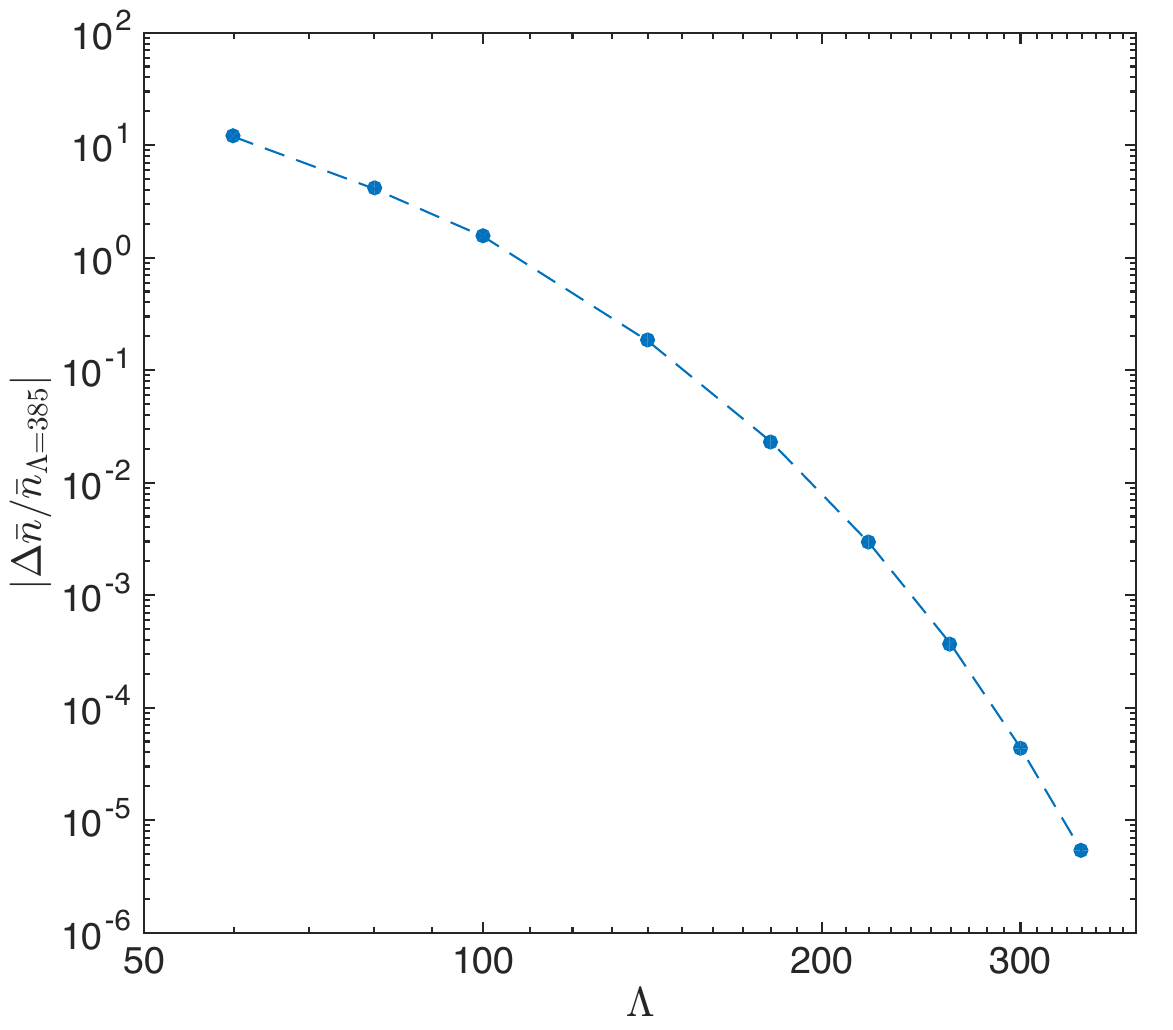}}
\subfloat{\includegraphics[width=7cm]{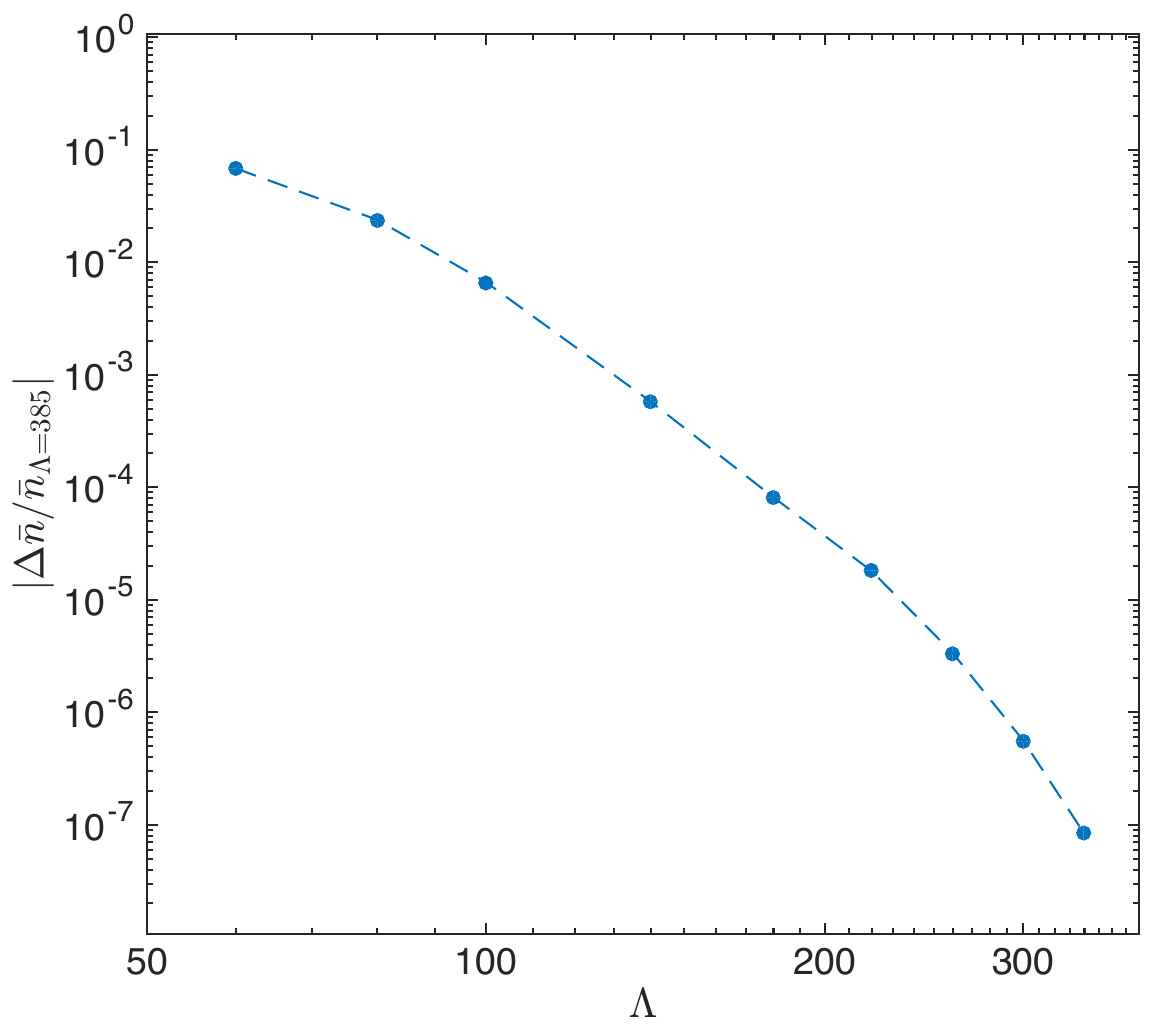}}
\caption{Error in the numerical result for the scattering phase $\phi(E)$ in the toy supermembrane model due to finite truncation energy $\Lambda$ (compared to the result at high truncation energy $\Lambda=385$),  at energies $E\simeq1.4$ (left) and $E\simeq102.5$ (right), with IR cutoff $L=20$.}
\label{fig:toyerr}
\end{figure}

In the $U=+1$ sector, the S-matrix has two eigenvalues, $\pm e^{i\phi(E)}$. This corresponds to a special case of (\ref{eq:aqc2}) with $k=2$, and we can extract $\phi(E)$ from the sequence of energy levels $\{E_n\}$ via
\ie
\phi(E_n) = \pi n - 2L \sqrt{E_n}.
\fe

\begin{figure}
\centering
\includegraphics[width=8.5cm]{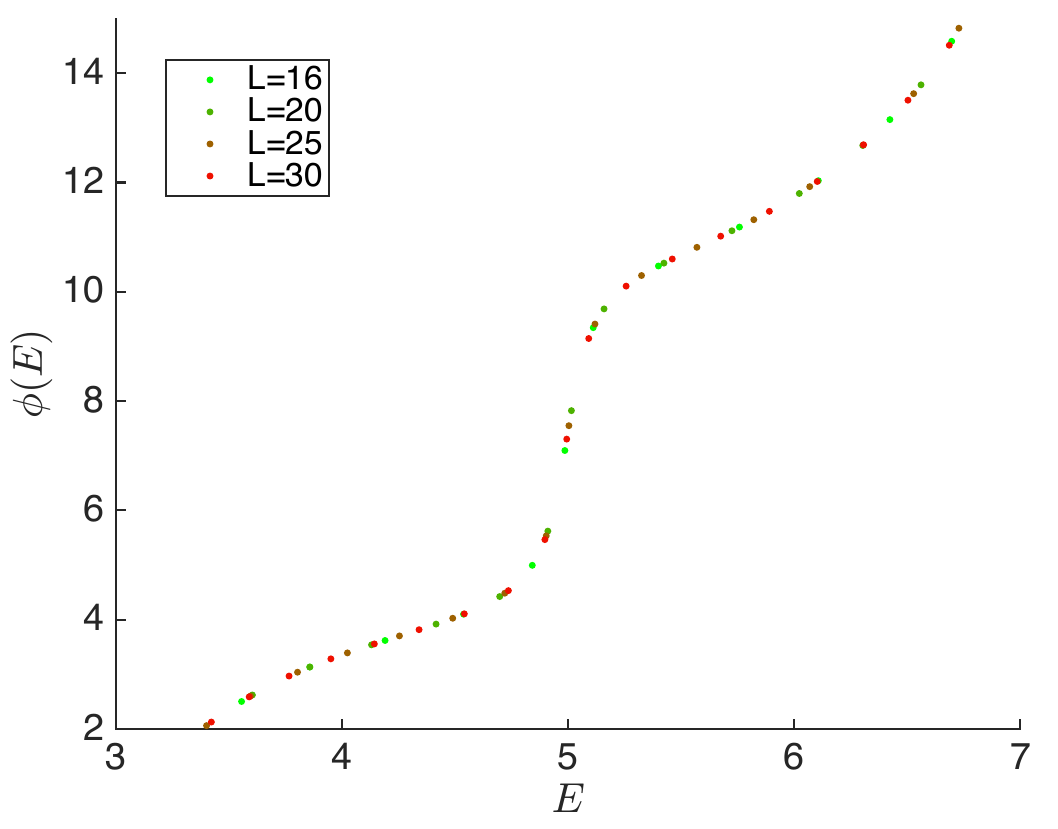}
\caption{The spectral sets with $L=16,20,25,30$ ($\Lambda=380$) lie on a single smooth curve. The abrupt increase of the scattering phase by $2\pi$ around $E\simeq5$ signals a metastable state. }
\bigskip
\bigskip
\centering
\includegraphics[width=10.8cm]{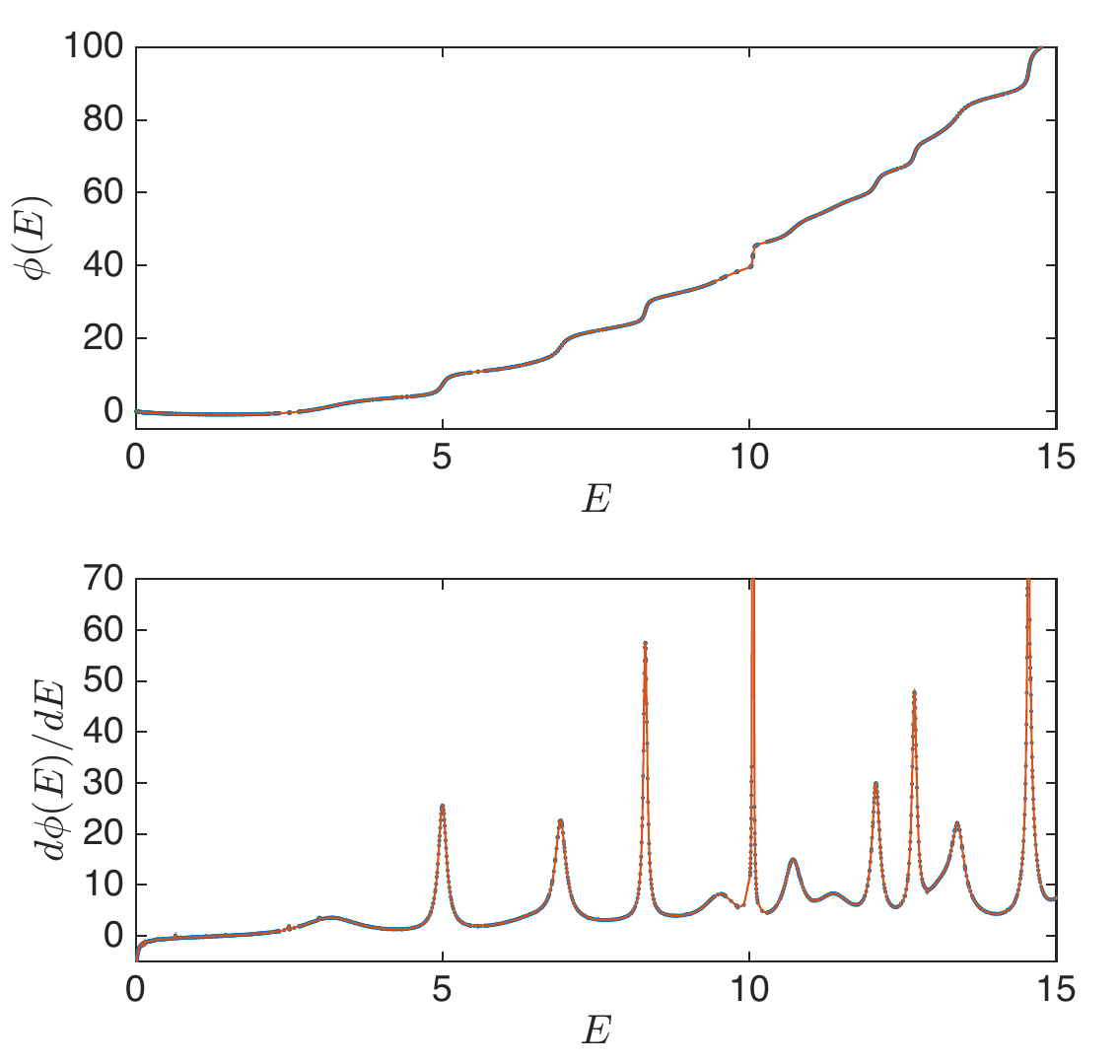}
\caption{Numerical results of $\phi(E)$ and $d\phi(E)/dE$ for the toy supermembrane model. In the first plot we have combined spectral sets with $L=16,17,\ldots,30$. We have used a locally weighted linear regression to smooth out the second plot.}
\label{fig:toyres}
\end{figure}

In Figure \ref{fig:toywavefns} we plot the density of the wave functions of a scattering state at generic energy and of a metastable state with very small decay width. The latter expectedly has localized support near the origin. 
The convergence of our numerical results with the truncation energy is exponentially fast, as confirmed in Figure \ref{fig:toyerr}. 

\begin{figure}[h]
\centering
\includegraphics[width=9cm]{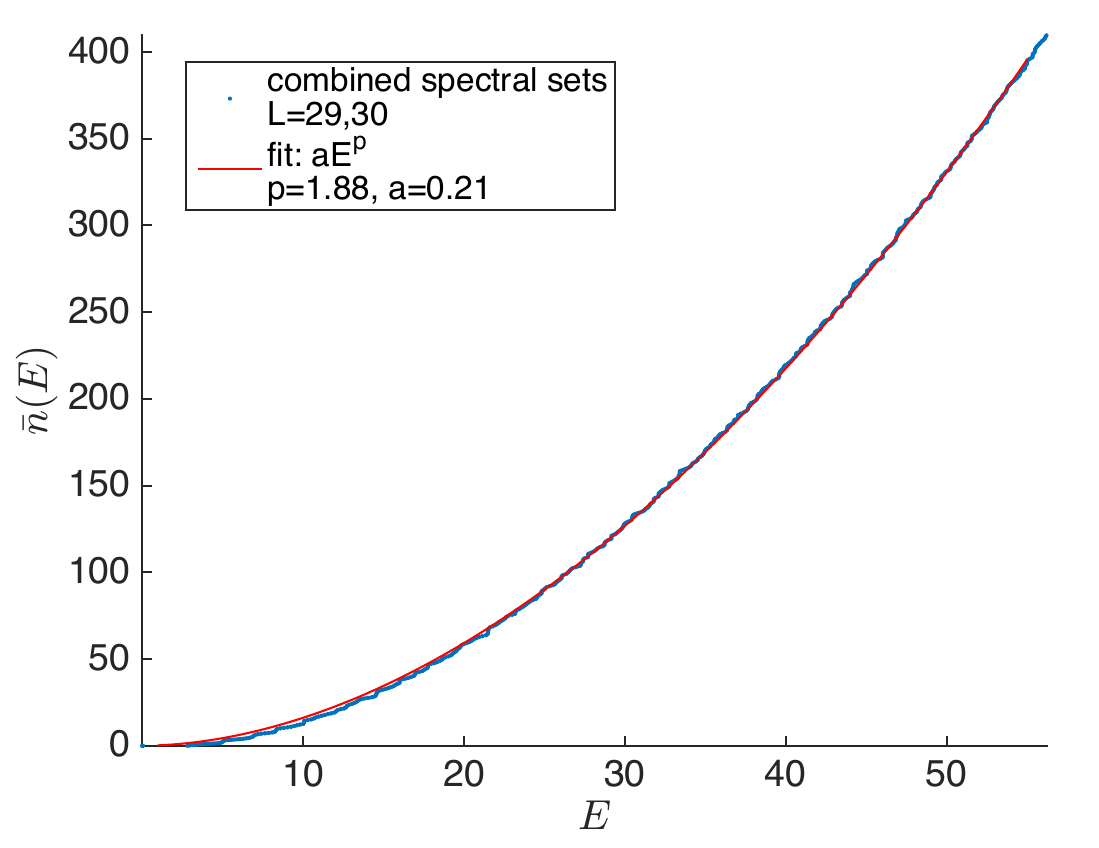}
\caption{High-energy scaling of $\bar{n}(E)$ for the toy supermembrane model in the $U=+1$ sector. We have combined spectral sets with $L=29$ and $30$, with truncation energy $\Lambda=380$. Up to energies where the Hamiltonian truncation approximation is reliable, we observe a power law $E^{p}$ with $p\approx 1.88$.}
\label{fig:toyreshigh}
\end{figure}

Our numerical results of scattering phase are shown in Figure \ref{fig:toyres} and \ref{fig:toyreshigh}. Note importantly that while the scattering phase $\phi(E)$ is a priori defined modulo $2\pi$, an abrupt increase of the phase by $2\pi$ as computed from the ``renormalized number of states" in energy is physical and signals a metastable or stable bound state. In principle, the metastable states can be detected from poles in the analytically continued S-matrix at complex values of $E$ with negative imaginary parts. When a pole is close to the real axis, it can be detected as a Breit-Wigner peak in $d\phi(E)/dE$.

Going to higher energies is more demanding numerically, as it requires increasing the truncation energy $\Lambda$ as well as increasing the IR cutoff $L$. In the energy range where our numerical results are reliable, we find a curious scaling behavior of the renormalized number of states, $\bar n(E)\sim E^{1.88}$ at high energies. Note that a naive semi-classical quantization would suggest a lower bound $\sim E^{3\over 2}$, which is also the result one would obtain using a 1-loop truncated Schwinger-Dyson equation \cite{Kabat:1999hp}.

\subsection{Metastable states}

\begin{figure}
\centering
\subfloat{\includegraphics[width=8cm]{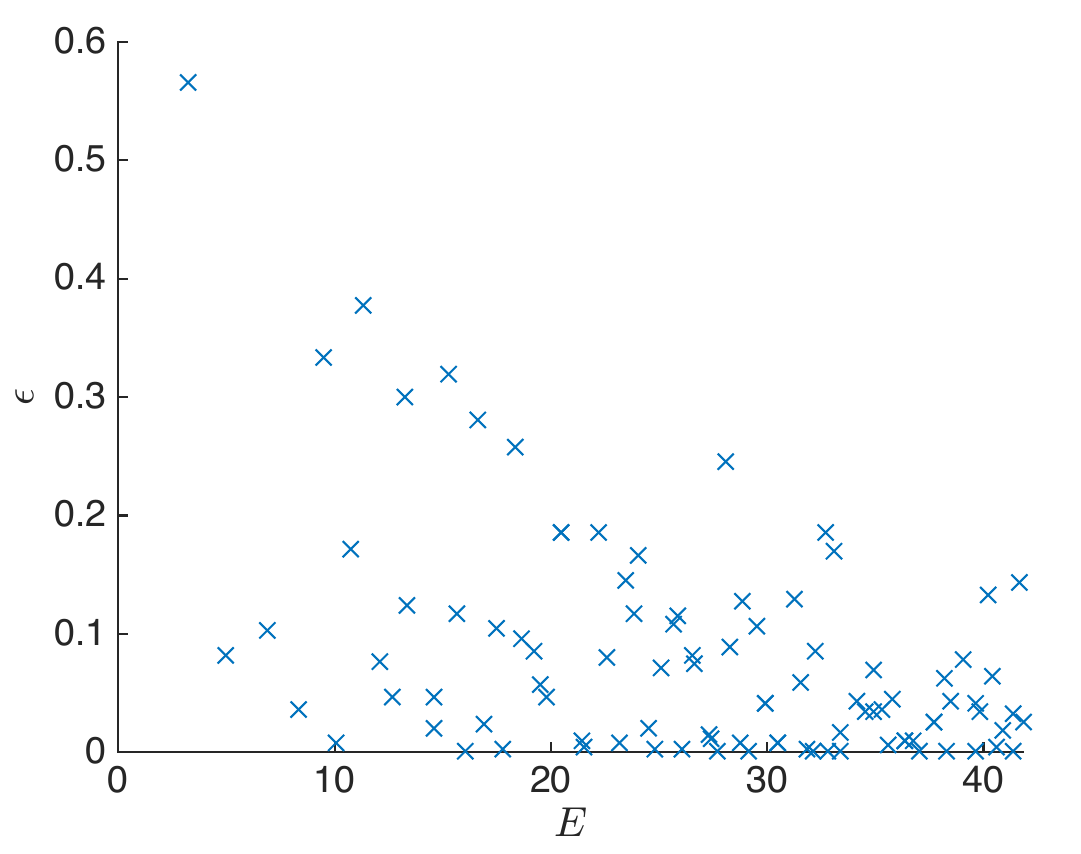}}
\subfloat{\includegraphics[width=8cm]{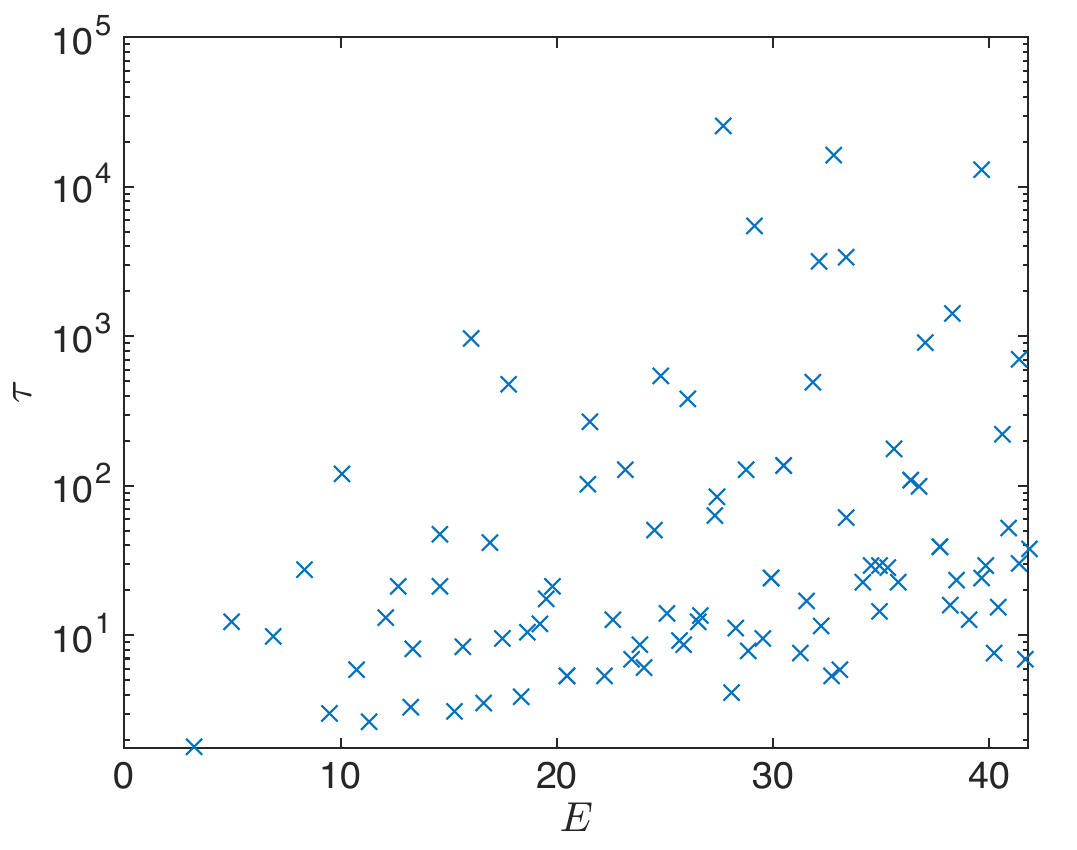}}
\caption{Breit-Wigner width $\epsilon$ (left) and lifetime (right) of the metastable states in the toy supermembrane model.}
\label{fig:BW}
\bigskip
\bigskip
\centering
\includegraphics[width=9cm]{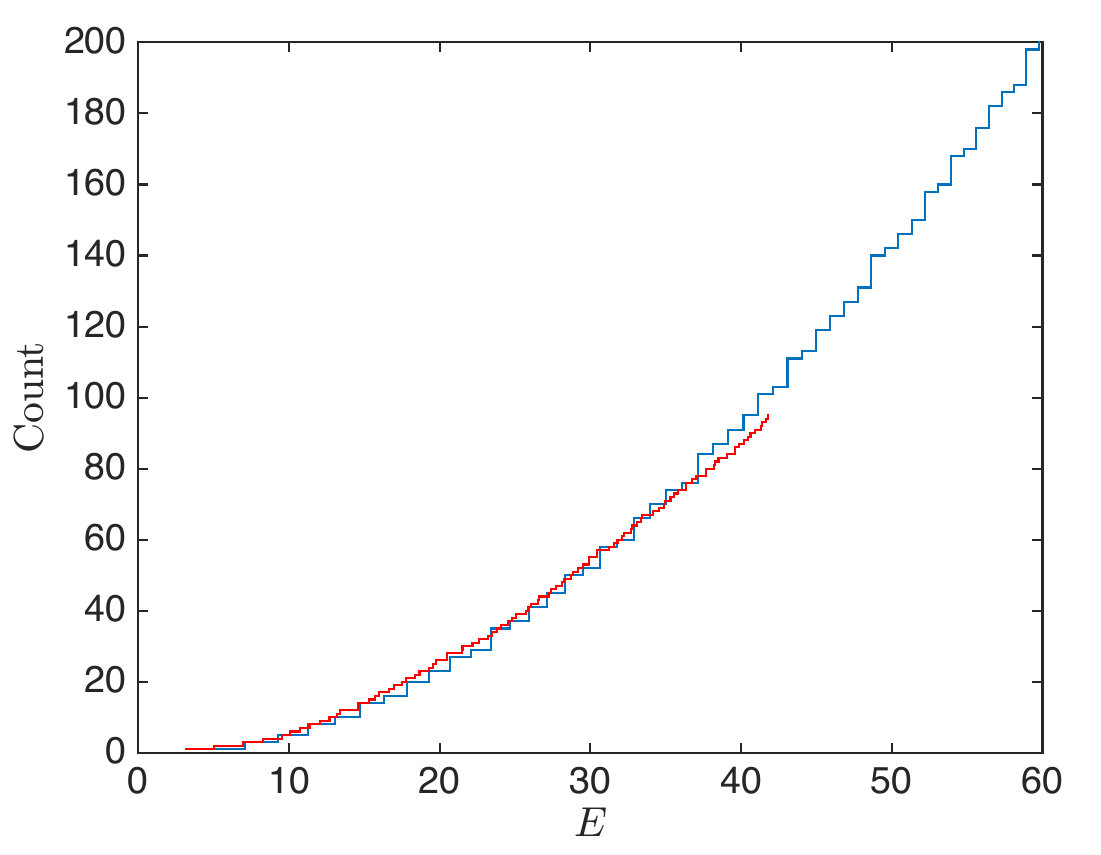}
\caption{The counting of metastable resonances (in the $U=+1$ sector, counted by 2-fold supermultiplets) up to a given energy $E$, in red, compared to the semi-classical estimate based on Born-Oppenheimer approximation, in blue.}
\label{fig:metstates}
\end{figure}

The metastable resonance energies and widths can be read off by fitting the peaks of the $d\phi(E)/dE$ plot to a Breit-Wigner distribution (\ref{eq:BW}), with the results shown in Figure \ref{fig:BW}. They can be compared to semi-classical expectations based on Born-Oppenheimer approximation. Along the valley of the interaction potential at large $x$, oscillation modes in the $y$-direction are approximated by a harmonic oscillator of frequency $|x|$. A mode in the $n$-th excitation level of the $y$-oscillator is governed by the effective Hamiltonian
\ie
H_{\mathrm{eff}}=-\partial_x^2+(2n+1)|x|+\sigma_3 x
\label{eq:effH}
\fe
in the large $x$ region, where $n=0,1,2,\ldots$. One component of the $n=0$ wave function corresponds to the asymptotic scattering state. The other modes are subject to a confining effective potential $V(x) = 2m|x|$, $m=1,2,\ldots$, giving rise to potential metastable states.

We may try to estimate the metastable resonance energies and their degeneracies using semi-classical Bohr-Sommerfeld quantization for this effective Hamiltonian. The result is $E_{m,\ell}=\left(3\pi m\ell\right)^\frac{2}{3}$, where $\ell=1,2,\ldots$ is the oscillation period in the effective potential in units of $2\pi \hbar$, and there is a further 4-fold degeneracy due to the two components of the wave function and the two asymptotic regions of large positive and negative $x$. This rough estimate is compared  to the direct enumeration of resonance peaks in $d\phi(E)/dE$, as shown in Figure \ref{fig:metstates} (where only the $U=+1$ supermultiplets are counted), with qualitative agreement at moderate energies.\footnote{A priori, the description of a metastable state with energy $E_{m,\ell}$ based on the Born-Oppenheimer approximation treating $y$ as ``heavy" modes and $x$ as ``light" modes is valid only for $\ell\gg m$. The rough agreement with counting of actual metastable resonances suggests that such a picture may be extended to all range of $m$ and $\ell$, with the role of $x$ and $y$ reversed in the $m\gg \ell$ regime. }

\section{An ${\cal N}=4$ supersymmetric model}
\label{sqm}

Now we turn to the scattering problem in a simple but nontrivial ${\cal N}=4$ SQM, namely that of a $U(1)$ gauge multiplet coupled to a charged chiral multiplet \cite{Denef:2002ru, Anous:2015xah}, with vanishing Fayet-Iliopoulos parameter. Here we follow the convention of \cite{Anous:2015xah}. The Hilbert space consists of 16-component wave functions on $\mathbb{R}^3\times \mathbb{C}$, parameterized by coordinates $\vec x\in \mathbb{R}^3$ associated with the gauge multiplet, and $(z,\bar z)\in\mathbb{C}$ associated with the chiral multiplet. The internal degrees of freedom of the wave function comes from quantization of gauginos $\lambda_\A, \bar\lambda^\B$ from the gauge multiplet and fermions $\psi_\A, \bar\psi^\B$ from the chiral multiplet. The fermions obey anti-commutation relations
\ie
\{ \lambda_\A, \bar\lambda^\B \} = \{ \psi_\A, \bar\psi^\B \} = \delta_\A^\B.
\fe
There are four supercharges, given by
\ie
& Q_\A = - {\sqrt{2}} \left[ \epsilon_{\A\B} \partial_{\bar z} + z x^i (\sigma_i)_\A{}^\C \epsilon_{\B\C}\right] \partial_{\psi_\B} + i (\sigma_i)_\A{}^\B \partial_{x_i} \lambda_\B - i|z|^2 \lambda_\A,
\\
& \bar Q^\A = - {\sqrt{2}} \left[ \epsilon^{\A\B} \partial_{ z} - \bar z x^i (\sigma_i)_\C{}^\A \epsilon^{\B\C}\right] {\psi_\B} + i (\sigma_i)_\B{}^\A \partial_{x_i} \partial_{\lambda_\B} + i|z|^2 \partial_{\lambda_\A}.
\fe
Here our convention for the antisymmetric tensor is $\epsilon_{\A\B}=-\epsilon^{\A\B}$, $\epsilon^{12}=1$. 
The supersymmetry algebra takes the form
\ie
& \{ Q_\A, Q_\B\} = 0 = \{\bar Q^\A, \bar Q^\B \},
\\
& \{ Q_\A, \bar Q^\B \} = 2 (\delta_\A{}^\B H - x^i (\sigma_i)_\A{}^\B G),
\fe
where $G$ is the $U(1)$ gauge rotation generator
\ie
G = \bar z \partial_{\bar z} - z \partial_z - \psi_\A  \partial_{\psi_\A} + 1,
\fe
and the Hamiltonian $H$ is given by
\ie
H = - \partial_z \partial_{\bar z} - {1\over 2}\partial_{x_i}^2 + {1\over 2} |z|^4 + x^2 |z|^2 - x^i (\sigma_i)_\A{}^\B \psi_\B \partial_{\psi_\A} + i \sqrt{2} z \epsilon_{\A\B} \partial_{\psi_\A} \partial_{\lambda_\B} - i \sqrt{2} \bar z \epsilon^{\A\B} \lambda_\A \psi_\B.
\fe
The wave functions are restricted to be invariant under $G$.

\subsection{Supermultiplets}

The model admits the symmetry $SU(2)_J\times U(1)_R$, where the $SU(2)_J$ rotates the $\mathbb{R}^3$ as well as the fermions, whereas the $U(1)_R$ rotates the fermions only. Their generators are
\ie
& \vec J = -i \vec x\times \vec\nabla_x + {1\over 2} \bar\psi \vec\sigma \psi + {1\over 2} \lambda\vec\sigma\lambda,
\\
& R = \bar\lambda\lambda - \bar\psi \psi.
\fe
The supercharges $Q_\A$ transforms as a doublet under $SU(2)_J$ and carry R-charge $-1$. Likewise, $\overline Q^\A$ carry R-charge $+1$. There is also a $\mathbb{Z}_2$ symmetry taking $R$ to $-R$.

The Hilbert space splits into sectors of R-charge $R=0$, $\pm1$, and $\pm2$. The $R=2$, $SU(2)$ spin $j$ states are necessarily annihilated by $\bar Q^\A$, and form a multiplet with $R=1$ spin $j\pm {1\over 2}$ and $R=0$ spin $j$ states. The $R=2$ sector does not admit scattering states, and thus they belong to supermultiplets that consist of only normalizable energy eigenstates, which we refer to as {\it $R=2$ multiplets}. Likewise, there is another multiplet consisting of stable $R=-2$ and $R=0$ states of spin $j$, and $R=-1$ states of spin $j\pm {1\over 2}$.

All scattering states must lie in supermultiplets that consist of $R=\pm 1$ states of spin $j$ and $R=0$ states of spin $j\pm {1\over 2}$. We refer to these supermultiplets as {\it $R=1$ multiplets}. The asymptotic region is given by the simultaneous limit of $r=|\vec x|\to \infty$ and $\tilde r=\sqrt{2}|z|\to 0$.

\begin{figure}[h]
\centering
\includegraphics[width=9cm]{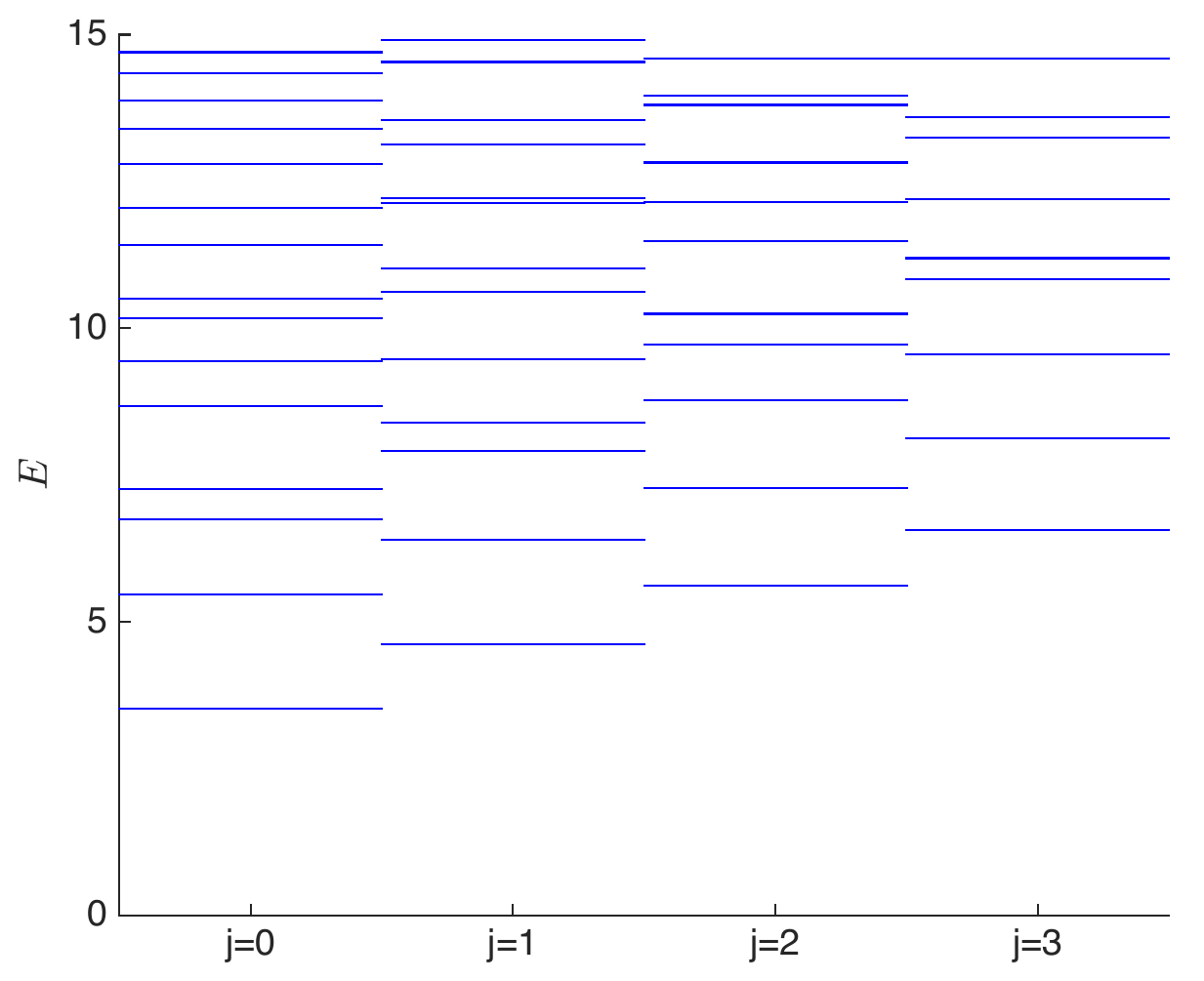}
\caption{Numerical spectra of bound states in the $R=2$ sector with $j=0,1,2,3$, computed with IR cutoffs $L=25$ and $\tilde L = 10$.}
\label{fig:R2evs}
\end{figure}

\subsection{Stable states from the $R=2$ multiplets}

It suffices to examine the $R=2$ sector wave function, whose $r, \tilde r$ dependence is governed by the reduced Hamiltonian 
\ie
H = H_0 + V,
\label{eq:R2H}
\fe
where \cite{Anous:2015xah}
\ie
& H_0 =  -{1\over 2r^2}\partial_r r^2 \partial_r - {1\over 2\tilde r} \partial_{\tilde r} \tilde r\partial_{\tilde r}+ {j(j+1)\over 2r^2} + {1\over 2\tilde r^2} ,
\\
& V = {\tilde r^4\over 8} + {r^2\tilde r^2\over 2} .
\fe
The matrix elements of the Hamiltonian are defined by the measure
\ie
\langle \psi' | H  | \psi\rangle = \int_0^\infty dr r^2 \int_0^\infty d\tilde r \tilde r \,\psi'(r,\tilde r)^\dagger H\psi(r,\tilde r).
\fe
The Hamiltonian (\ref{eq:R2H}) has a discrete spectrum and thus we can adopt the standard Rayleigh-Ritz method, imposing an IR cutoff $r<L$ and $\tilde r<\tilde L$ and demanding that the wave function vanishes at $r=L$ and at $\tilde r=\tilde L$. Note that the wave function need not vanish at $r=0$ or at $\tilde r=0$. 

Denote by 
\ie{}
& f_{j,n}(r) = {a_{j,n} \over \sqrt{r}}J_j(\sqrt{2\omega_{j,n}} r) 
\fe
a basis of orthonormal functions on the spherical box of unit radius. The frequencies $\omega_n$ are determined by the boundary condition $J_j(\sqrt{2\omega_{j,n}})=0$. A basis of wave functions that diagonalize $H_0$ are
\ie
\psi_{n,m}(r,\tilde r) = (L\tilde L)^{-{3\over 2}} f_{j+{1\over 2},n}(r/L) \sqrt{\tilde r} f_{1,m}(\tilde r/\tilde L).
\fe
Note that here $j$ is an integer, and so the $r$-dependence of the basis wave functions can be expressed in terms of elementary functions, while the $\tilde r$-dependence is expressed through the Bessel function $J_1$.

We then diagonalize the matrix
\ie\label{normh}
\langle n,m| H|k,\ell\rangle = \delta_{nm}\delta_{k\ell} \left( {\omega_{j+{1\over 2},n}\over L^2} + {\omega_{1,m}\over \tilde L^2} \right) + \int_0^L dr r^2 \int_0^{\tilde L} d\tilde r \tilde r \, \psi_{n,m}^\dagger V \psi_{k,\ell},
\fe
with a suitable level truncation on $n,m,k,\ell$. Numerical results are shown in Figure \ref{fig:R2evs}.

\begin{figure}
\centering
\includegraphics[width=9.5cm]{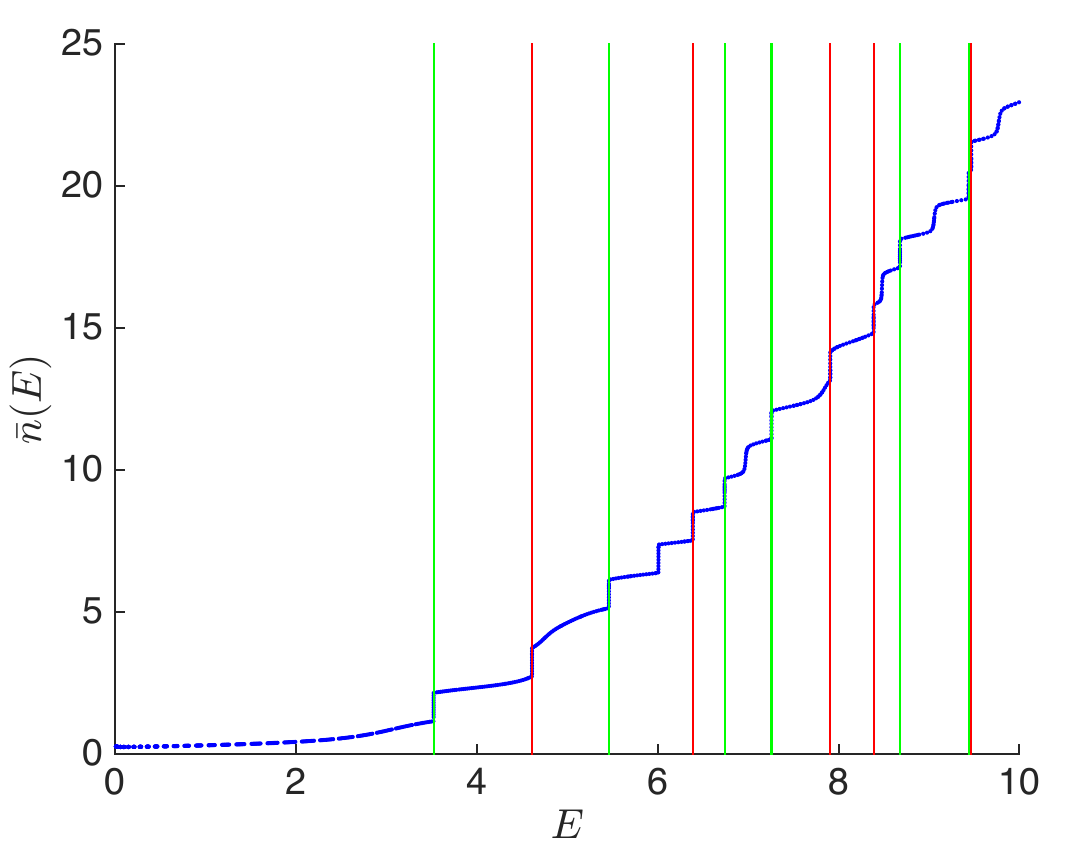}
\caption{Numerical results of $\overline{n}(E)$ in the $R=1,j=1/2$ sector. We have combined spectral sets with $L=30,30.1,30.2,\ldots,31$, $\tilde{L}=10$, and truncation energy $\Lambda=250$. Vertical lines indicate a stable state in the $R=2$, $j=0$ (green) and $j=1$ (red) sectors.}
\bigskip
\centering
\subfloat{\includegraphics[width=5.5cm]{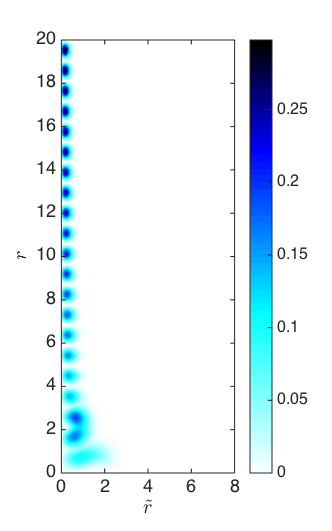}}
\subfloat{\includegraphics[width=5.65cm]{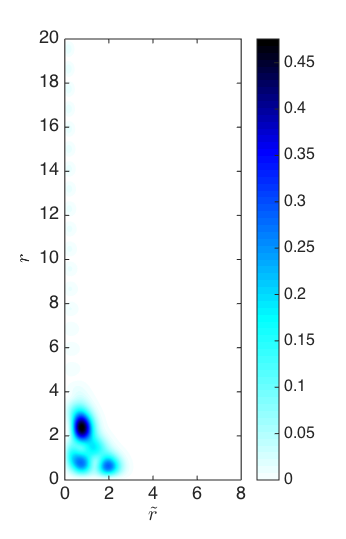}}
\caption{Probability density $\tilde{r}r^2|\Psi|^2$ of a generic scattering wave function at $E=5.59$ (left) and of a metastable state at $E=6.01$ (right), in the $R=1, j=1/2$ multiplets. Here we have taken the IR cutoffs $L=20$, $\tilde{L}=8$, and truncation energy $\Lambda=220$.}
\label{fig:N4wavef}
\end{figure}

\subsection{The $R=1$ sector}

To compute the S-matrix of scattering states and metastable resonances, we will study the $R=1$, spin $j$ sector using the Hamiltonian truncation method. In the spectrum we will find scattering states that are in a supermultiplet with $R=0$ and $R=-1$ states, as well as a discrete set of stable states that are in a supermultiplet with $R=0$ and $R=2$ or $R=-2$ states. By comparison with the spectrum of $R=2$ sector, we will be able to remove the stable states from the spectrum of the $R=1$ sector, and identify the density of scattering states (subtracting IR divergent contribution) and thereby their scattering matrix.

\begin{figure}
\centering
\includegraphics[width=11cm]{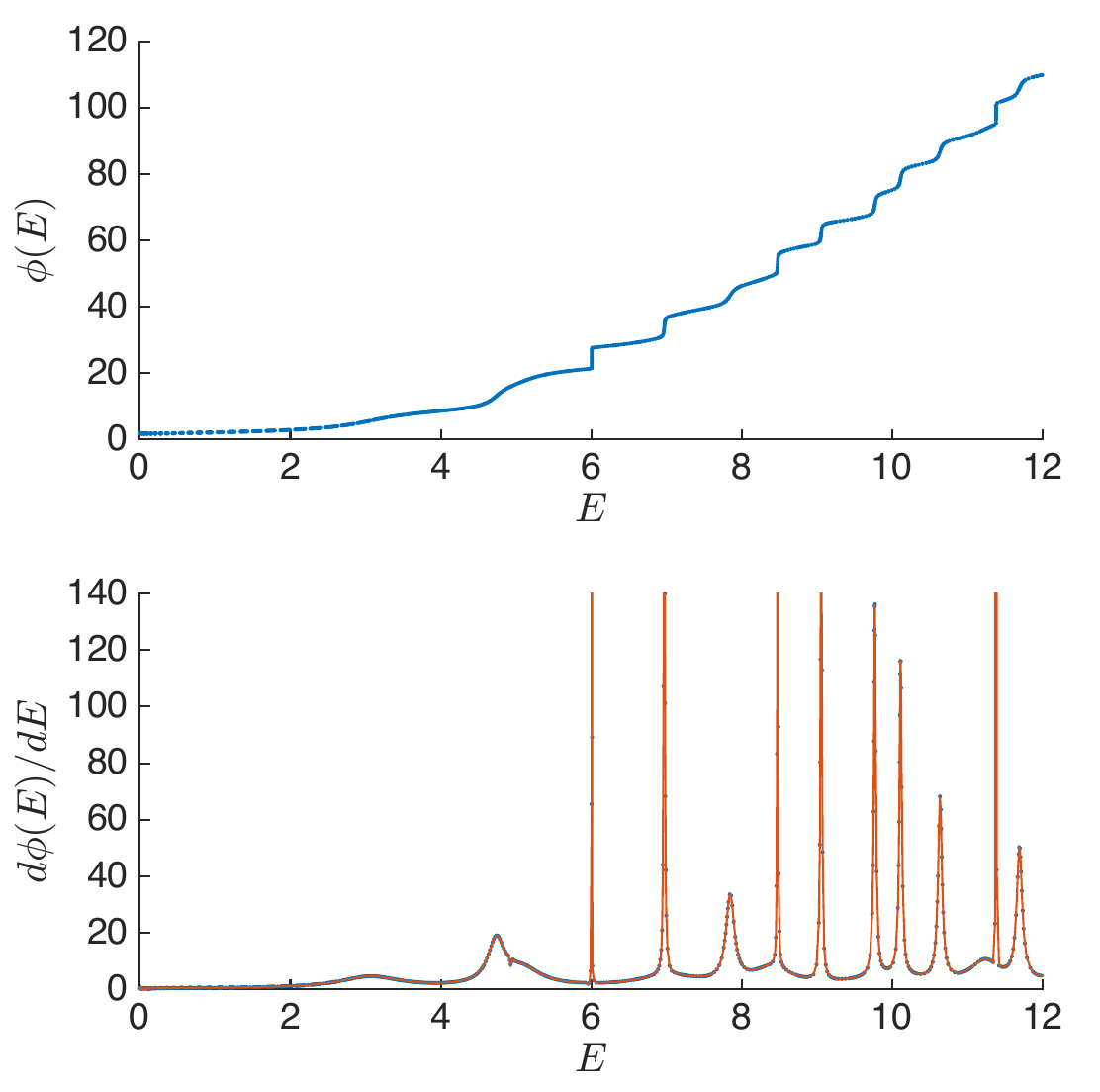}
\caption{Numerical results of $\phi(E)$ and $d\phi(E)/dE$ for the $R=1$ multiplets with $j=1/2$, after removing the stable states in the $R=2$ multiplets with $j=0,1$ from the $R=1$ sector. We have combined spectral sets with IR cutoff $L=30,30.1,30.2,\ldots,31$, $\tilde{L}=10$, and truncation energy $\Lambda=250$. We have used a locally weighted linear regression to smooth out the second plot.}\label{Ronewidth}
\bigskip
\subfloat{\includegraphics[width=8cm]{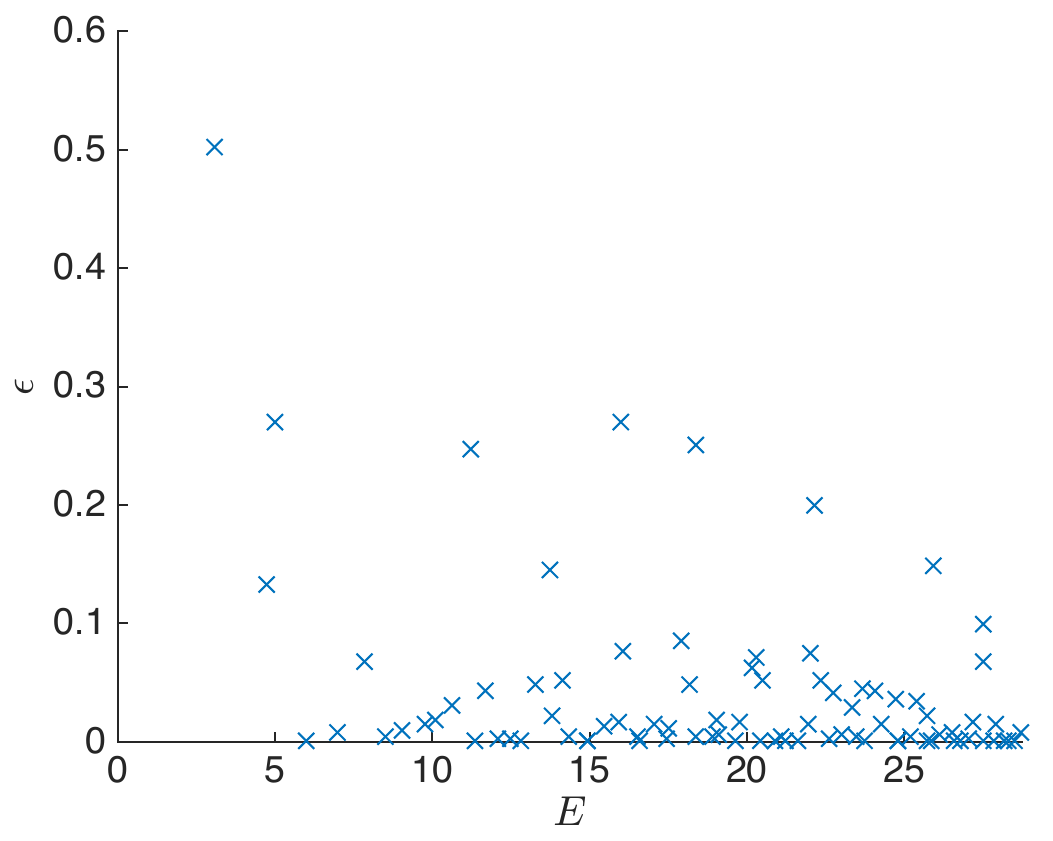}}
\subfloat{\includegraphics[width=8cm]{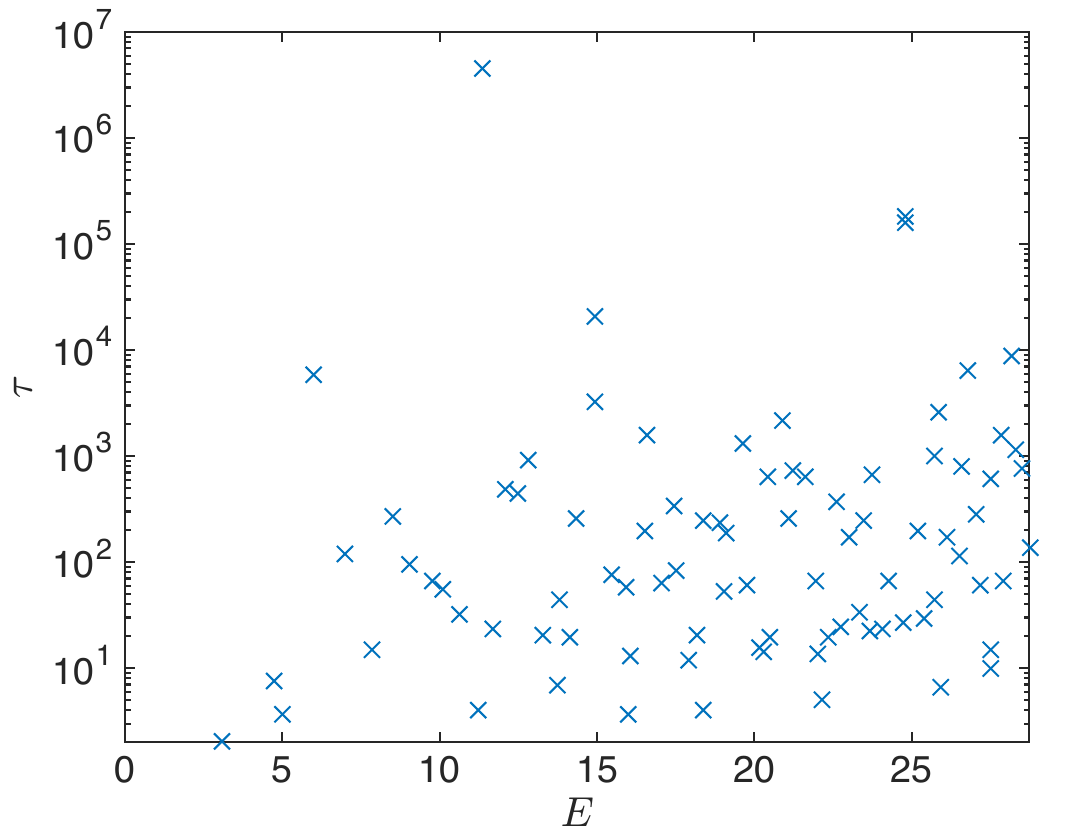}}
\caption{Breit-Wigner width (left) and lifetime (right) of the metastable states in the $R=1, j=1/2$ multiplets.}
\label{fig:BWn4}
\end{figure}

The reduced Hamiltonian of the $R=\pm 1$ sector can be written in the form
\ie
& H_{j,|R|=1} = H_0 + V, 
\fe
where \cite{Anous:2015xah}
\ie{}
& H_0 =  -{1\over 2r^2}\partial_r r^2 \partial_r - {1\over 2\tilde r} \partial_{\tilde r} \tilde r\partial_{\tilde r} + \begin{pmatrix} {4j^2-1\over 8r^2} & 0 & 0 & 0 \\ 0 & {4j(j+2)+3\over 8 r^2} & 0 & 0 \\ 0 & 0 & {4j^2-1\over 8r^2} + {1\over 2\tilde r^2} & 0 \\ 0 & 0 & 0 & {4j(j+2)+3\over 8 r^2}+{1\over 2\tilde r^2} \end{pmatrix},
\\
& V =  {\tilde r^4\over 8} + {r^2\tilde r^2\over 2} + \begin{pmatrix} ~0 ~& ~r~ & ~\tilde r~ & ~0~ \\ r & 0 & 0 & \tilde r \\ \tilde r & 0 & 0 & 0 \\ 0 & \tilde r & 0 & 0 \end{pmatrix} .
\fe
Here $j$ takes positive half-integer values.
The Hilbert space is spanned by 4-component wave functions $\psi(r,\tilde r)$ supported in the domain $r,\tilde r\geq 0$, subject to the norm (\ref{normh}).

A basis of wave functions that diagonalize $H_0$ are
\ie{}
(\psi^{s}_{n,m})_t = (L\tilde L)^{-{3\over 2}} f_{j+\delta_{s2}+\delta_{s4},n}(r/L) \sqrt{\tilde r} f_{\delta_{s3}+\delta_{s4},m}(\tilde r/\tilde L) \delta_{st},
~~~~s,t=1,2,3,4.
\fe

%

We will then diagonalize the matrix
\ie{}
&\langle n,m,s| H_{j,|R|=1} |k,\ell,t\rangle \\
&= \delta_{nm} \delta_{k\ell} \delta_{st}  \left({\omega_{j+\delta_{s2}+\delta_{s4},n}\over L^2} + {\omega_{\delta_{s3}+\delta_{s4},m}\over \tilde L^2}\right)
+ \int_0^L dr r^2 \int_0^{\tilde L} d\tilde r \tilde r \, \psi_{n,m}^s V \psi_{k,\ell}^t
\label{eq:R1trH}
\fe
with a suitable level truncation on $n,m,k,\ell$. 

Note that despite that the wave function has 4 components, there is only one scattering state at a given energy $E$. Therefore, the S-matrix is determined by a single scattering phase. The asymptotic wave function takes the form
\ie
\psi(r,\tilde r) \sim \begin{pmatrix} 1\\ -1 \\0 \\0 \end{pmatrix} \left[ e^{-i \sqrt{2E} r} + e^{i\phi(E)} e^{i \sqrt{2E} r} \right] {e^{- r \tilde r^2/2}\over r} .
\fe
The number of states only has a linear divergence in $L$. After subtracting off the IR divergence, we obtain the renormalized number of states
\ie
\overline{n}(E_n) = n+{1\over 2} - { L \over\pi} \sqrt{2E_n}.
\fe
In Figure \ref{fig:N4wavef} we plotted the density of the wave functions in the $(\tilde r, r)$ plane of a scattering state at a generic energy and of a metastable state with very small decay width.

\subsection{Scattering states from the $R=1$ multiplets}

The stable states in the $R=2$ sector with spin $j$ are in the same supermultiplet as $R=1$ states with spin $j\pm {1\over 2}$. Therefore, the spectrum of scattering states of $R=1$ and spin $j$ can be obtained by removing from the $R=1$ spectrum stable states in the same supermultiplet as $R=2$, spin $j\pm{1\over 2}$ states. The resulting scattering phase $\phi(E)$, and the metastable resonances from $d\phi(E)/dE$, are shown in Figure \ref{Ronewidth}. Figure \ref{fig:BWn4} shows the metastable resonance energies and widths obtained by fitting to a Breit-Wigner distribution.

\begin{figure}[h]
\centering
\includegraphics[width=12cm]{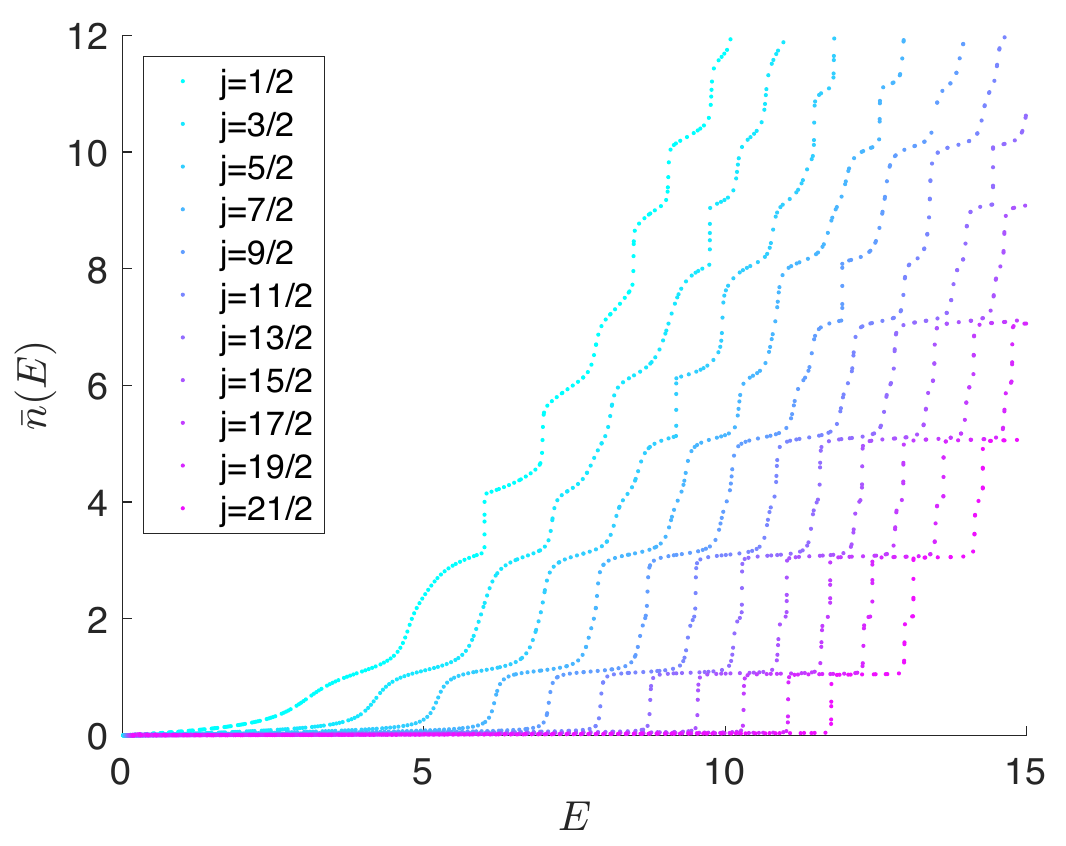}
\caption{Numerical results of $\bar n(E)$ for the $R=1$ multiplets with $j={1\over 2},{3\over 2},\ldots,{21\over 2}$ as indicated in the legend, after removing the stable states in $R=2$ multiplets from the $R=1$ sector, and correcting a finite $L$ effect due to a rotational effective potential $j(j+1)/2r^2$ that affects low energy scattering states. Here we have taken the IR cutoffs $L=25$, $\tilde{L}=5$ and truncation energy $\Lambda=200$.}\label{fig:highjps}
\end{figure}

As $j$ increases, the effective potential pushes the wave function away from the origin, and it becomes important to correct for the finite $L$ effects in order to obtain accurate scattering phases at low energies. This is achieved by comparison with the free Hamiltonian with an effective rotational potential ${j(j+1)\over 2r^2}$ at the same IR cutoff $L$, and subtract off the contribution to $\bar n(E)$ of this system (which admits a trivial S-matrix). The scattering phases (up to constant shifts) for $j$ up to ${21\over 2}$ are shown in Figure \ref{fig:highjps}. The results are in agreement with the Coulomb branch effective Hamiltonian \cite{Pioline:2015wza} which is valid in the limit of fixed $E$ and large $j$, that gives a trivial scattering phase (when the FI parameter is set to zero). The widths of the metastable resonances generally become narrower as $j$ increases, as the wave functions are pushed to large values of $r$, and the excited modes in $\tilde r$ direction become more stable against decaying to the asymptotic states.

\section{Discussion}
\label{discussion}

While the Rayleigh-Ritz method is very general and can be applied to any quantum system, supersymmetry provides interesting classes of models with flat directions and nontrivial spectra of metastable states, and in particular precise holographic models that allow for accessing the semi-classical gravity regime in the bulk. To develop tools for analyzing the real time dynamics of strongly coupled SQM and the unitary evolution of black hole microstates is the main motivation for this work.

In applying the Rayleigh-Ritz method to the numerical study of the S-matrix, two key observations were made in this paper. Firstly, we do not need to take the IR cutoff $L$ to be excessively large; on the other hand, it is extremely important to sample over a set of moderately large $L$ values, so that the spectral set $\{(E_n, \overline n)\}$ trace out the curves of scattering phases. This is crucial for determining the metastable resonances and their widths to high accuracy. Secondly, by varying the IR cutoffs in various asymptotic regions, we can extract not just a single scattering phase but the entire S-matrix. The second observation is not necessary for the models considered in this paper, but will be useful for more general models that have several asymptotic regions unrelated by symmetries, such as SQMs that admit noncompact Higgs and Coulomb branches.

Ultimately, we would like to implement our method in the BFSS matrix quantum mechanics \cite{Banks:1996vh} and extract the spectrum of black hole microstates in the holographic dual \cite{Banks:1997hz, Balasubramanian:1997kd, Susskind:1998vk, Polchinski:1999ry}. This is substantially more complicated than the toy supermembrane model and the ${\cal N}=4$ SQM considered here, but they share a number of common features. While in principle the large $N$ limit is required for accessing the semi-classical gravity regime in the bulk, there are indications that this may not be entirely necessary \cite{Anagnostopoulos:2007fw, Hanada:2009ne, Hanada:2016zxj}, and we may extract relevant physics even for $N=2$ or 3. The $SU(2)$ BFSS MQM involves a priori $2^{24}$ components of the wave function that depend on 27 bosonic coordinates. However, there are only 3 $SO(9)\times SU(2)$ invariants formed out of the bosonic coordinates, which will play the role of $r,\tilde r$ variables in the ${\cal N}=4$ SQM analyzed in this paper. Furthermore, the $2^{24}$ component fermionic wave function reduces to a few hundred irreducible representations of $SO(9)\times SU(2)$ \cite{Hoppe:2008uc, Michishita:2010rx}, and the direct diagonalization of a truncated Hamiltonian could be manageable when restricted to a sector with fixed $SO(9)$ angular momentum. We hope to report on this in the near future.

\section*{Acknowledgements}

We would like to thank Balt van Rees and Slava Rychkov for enlightening conversations, and  Shu-Heng Shao for comments on a preliminary draft. XY would like to thank Korea Institute for Advanced Study and Stony Brook University for their hospitality during the course of this work. This work is supported by a Simons Investigator Award from the Simons Foundation, and in part by DOE grant DE-FG02-91ER40654. VR was supported by the National Science Foundation Graduate Research Fellowship under Grant No. DGE1144152. The numerical computations in this work are performed on the Odyssey cluster supported by the FAS Division of Science, Research Computing Group at Harvard University.

\appendix

\section{Some details of numerical implementation}

There are two main steps in the numerical implementation of the Rayleigh-Ritz method, building the Hamiltonian matrix and diagonalizing it. In this section, we will describe the approach used for the ${\cal N}=4$ supersymmetric quantum mechanics model. 

\begin{figure}[h]
\centering
\subfloat{\includegraphics[width=7.5cm]{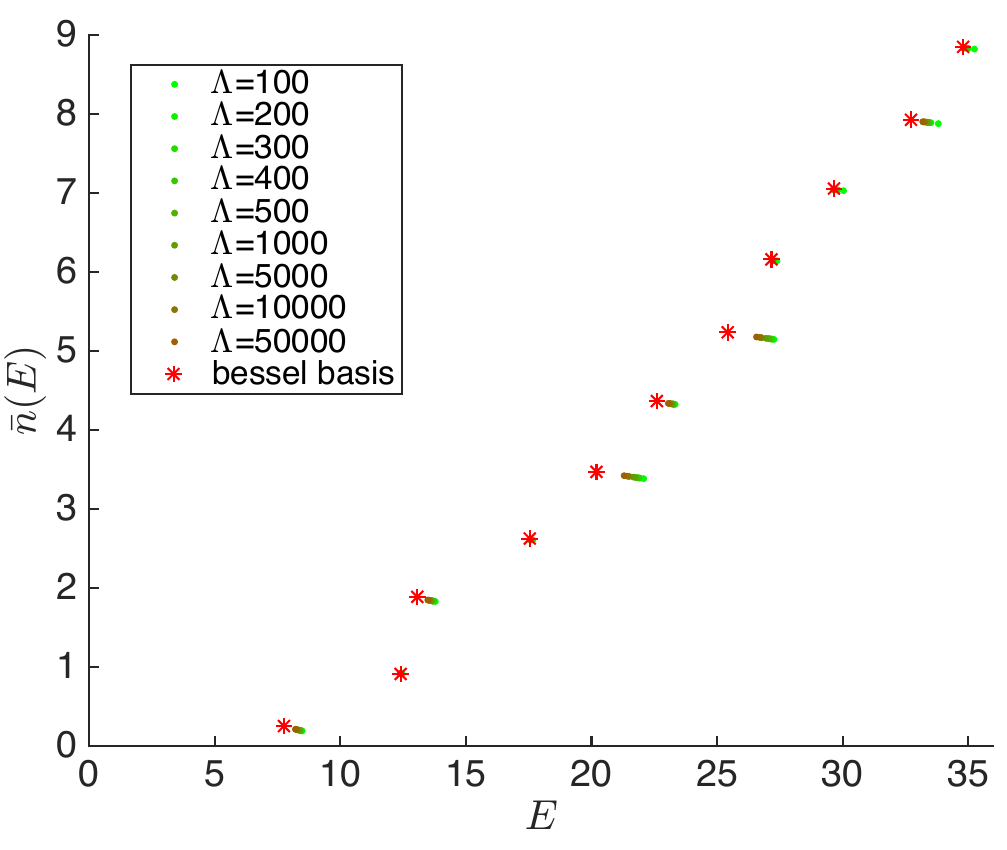}}
\subfloat{\includegraphics[width=7.5cm]{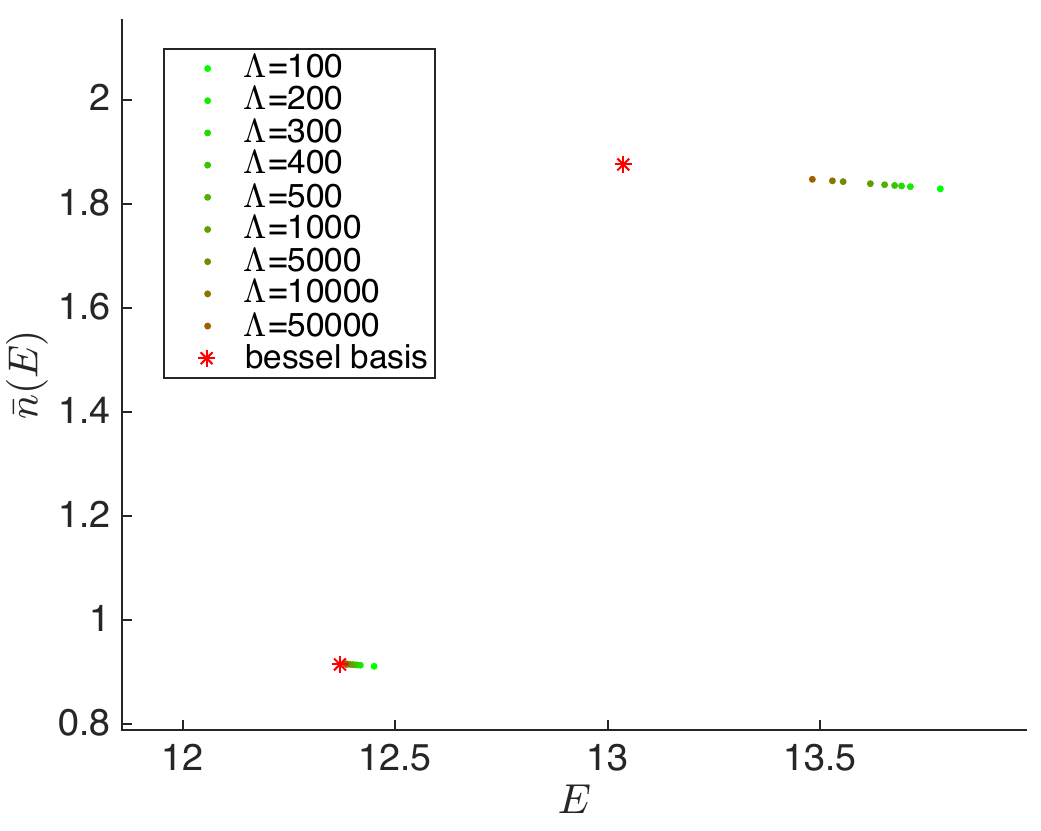}}
\caption{Energy levels with IR cutoff $L=1$ and $\tilde L=1$, computed with the Bessel function basis versus a simpler basis $\psi_{n,m}(r,\tilde{r})={2\over r\sqrt{\tilde r}\sqrt{L\tilde L}  }\sin({n\pi r\over L})\sin({m\pi\tilde{r}\over \tilde{L}})$ adapted to a free Hamiltonian that does not include the rotational potential. While the spectrum computed using the Bessel basis converges exponentially fast, the results computed using the simpler basis have very slow convergence with truncation energy $\Lambda$ (and the convergence becomes worse with increasing $L$).}
\label{fig:badbasis}
\end{figure}

The first step is to compute the matrix elements of the truncated Hamiltonian as in (\ref{eq:R1trH}). 
As discussed for the 1D models in section 3.1, it is important to split the Hamiltonian such that the interaction term is smooth in $r$ and $\tilde r$. One could use a basis that does not leave a smooth interaction term, but the convergence in truncation energy becomes slower, requiring significantly more memory resources and time (especially in the diagonalization). To avoid this, we used a basis of Bessel functions which despite requiring numerical integrations for the matrix elements of the Hamiltonian, leads to exponential convergence in truncation energy. We illustrate this with a sample calculation at small IR cutoff lengths $L$ and $\tilde L$ in Figure \ref{fig:badbasis}.

We parallel processed the computation of matrix elements using the `parfor' function of MATLAB's Parallel Computing Toolbox. A typical run of our code used a pool of 30 workers to distribute the process. 

The second step is to diagonalize the matrix. For that end we used MATLAB's function eig() that automatically implements multithreading of available cores, and hence no explicit parallelization is needed. A few typical runs and the parameters of the numerics are shown in Table 1.

%
   
\begin{center} 
  \begin{tabular}{c |  c | c | c | c |}
    \cline{2-5}
    & \multirow{2}{*}{$\Lambda$} & \multicolumn{2}{ c| }{Time} & \multirow{2}{*}{Matrix size} \\ \cline{3-4}
    & & \multicolumn{1}{c|}{Matrix elements}  & \multicolumn{1}{c|}{Diagonalization} & \\ \hline
   \multicolumn{1}{|c|}{\parbox[t]{2mm}{{\multirow{4}{*}{\rotatebox[origin=c]{90}{$\cal{N}$$=4$}}}}}& 100 & \multicolumn{1}{c|}{23 min} & 2 min & $4000$ \\ 
      \multicolumn{1}{|c|}{}& 200 & \multicolumn{1}{c|}{1.3 hr} & 8 min & $8000$ \\
      \multicolumn{1}{|c|}{}& 300 & \multicolumn{1}{c|}{3.5 hr} & 23 min & $25000$ \\
      \multicolumn{1}{|c|}{}& 400 & \multicolumn{1}{c|}{8.9 hr} & 55 min & $33000$ \\ 
     \hline
    \multicolumn{1}{|c|}{\parbox[t]{2mm} {\multirow{3}{*}{\rotatebox[origin=c]{90}{Toy}}}} &100 & 75 sec & 9 min & $16000$ \\ 
    \multicolumn{1}{|c|}{}& 300 & 13 min & 3.9 hr & $48000$ \\
    \multicolumn{1}{|c|}{}& 500 & 56 min & 12.5 hr & $81000$ \\
    \hline
  \end{tabular}
\captionof{table}{Sample parameters of a typical run for the ${\cal N}=4$ model in the $R=1$, $j=1/2$ sector with $L=\tilde{L}=10$, and for the toy supermembrane model with $L=20$, computed with a pool of 30 workers.}
\end{center}




\bibliographystyle{JHEP}
\bibliography{QMdraft.bib}

\end{document}